\documentclass{article}
\usepackage{spconf,amsmath,graphicx}
\usepackage{url}
\usepackage{subfig}
\usepackage{xcolor}

\usepackage{booktabs, multirow} 
\usepackage{hyperref,xcolor}
\hypersetup{
    colorlinks=true,
    linkcolor=purple,
    filecolor=magenta,      
    urlcolor=purple,
}
\usepackage{cleveref}
\usepackage{soul}
\usepackage{amsfonts}
\usepackage{pifont}
\newcommand{\cmark}{\ding{51}}%
\newcommand{\xmark}{\ding{55}}%

\usepackage[
backend=biber,
style=ieee,
citestyle=numeric-comp,
maxbibnames=1,
maxcitenames=1,
doi=false,isbn=false,url=false,eprint=false
]{biblatex}

\defbibheading{bibliography}[\refname]{}

\DeclareSourcemap{
	\maps[datatype=bibtex, overwrite=true]{
		\map{
			\step[fieldsource=booktitle,
			match=\regexp{.*Interspeech.*},
			replace={Proc. Interspeech}]
			\step[fieldsource=journal,
			match=\regexp{.*INTERSPEECH.*},
			replace={Proc. Interspeech}]
			\step[fieldsource=booktitle,
			match=\regexp{.*ICASSP.*},
			replace={Proc. ICASSP}]
			\step[fieldsource=booktitle,
			match=\regexp{.*icassp_inpress.*},
			replace={Proc. ICASSP (in press)}]
			\step[fieldsource=booktitle,
			match=\regexp{.*Acoustics,.*Speech.*and.*Signal.*Processing.*},
			replace={Proc. ICASSP}]
			\step[fieldsource=booktitle,
			match=\regexp{.*International.*Conference.*on.*Learning.*Representations.*},
			replace={Proc. ICLR}]
			\step[fieldsource=booktitle,
			match=\regexp{.*International.*Conference.*on.*Computational.*Linguistics.*},
			replace={Proc. COLING}]
			\step[fieldsource=booktitle,
			match=\regexp{.*SIGdial.*Meeting.*on.*Discourse.*and.*Dialogue.*},
			replace={Proc. SIGDIAL}]
			\step[fieldsource=booktitle,
			match=\regexp{.*International.*Conference.*on.*Machine.*Learning.*},
			replace={Proc. ICML}]
			\step[fieldsource=booktitle,
			match=\regexp{.*North.*American.*Chapter.*of.*the.*Association.*for.*Computational.*Linguistics:.*Human.*Language.*Technologies.*},
			replace={Proc. NAACL}]
			\step[fieldsource=booktitle,
			match=\regexp{.*Empirical.*Methods.*in.*Natural.*Language.*Processing.*},
			replace={Proc. EMNLP}]
			\step[fieldsource=booktitle,
			match=\regexp{.*Association.*for.*Computational.*Linguistics.*},
			replace={Proc. ACL}]
			\step[fieldsource=booktitle,
			match=\regexp{.*Automatic.*Speech.*Recognition.*and.*Understanding.*},
			replace={Proc. ASRU}]
			\step[fieldsource=booktitle,
			match=\regexp{.*Spoken.*Language.*Technology.*},
			replace={Proc. SLT}]
			\step[fieldsource=booktitle,
			match=\regexp{.*Speech.*Synthesis.*Workshop.*},
			replace={Proc. SSW}]
			\step[fieldsource=booktitle,
			match=\regexp{.*workshop.*on.*speech.*synthesis.*},
			replace={Proc. SSW}]
			\step[fieldsource=booktitle,
			match=\regexp{.*Advances.*in.*neural.*information.*processing.*},
			replace={Proc. NeurIPS}]
			\step[fieldsource=booktitle,
			match=\regexp{.*Advances.*in.*Neural.*Information.*Processing.*},
			replace={Proc. NeurIPS}]
			\step[fieldsource=booktitle,
			match=\regexp{.*Workshop.*on.* Applications.* of.* Signal.*Processing.*to.*Audio.*and.*Acoustics.*},
			replace={Proc. WASPAA}]
			\step[fieldsource=publisher,
			match=\regexp{.+},
			replace={{}}]
			\step[fieldsource=month,
			match=\regexp{.+},
			replace={{}}]
			\step[fieldsource=location,
			match=\regexp{.+},
			replace={{}}]
			\step[fieldsource=address,
			match=\regexp{.+},
			replace={{}}]
			\step[fieldsource=organization,
			match=\regexp{.+},
			replace={{}}]
			\step[fieldsource=pages,
			match=\regexp{.+},
			replace={{}}]
			\step[fieldsource=volume,
			match=\regexp{.+},
			replace={{}}]
			\step[fieldsource=number,
			match=\regexp{.+},
			replace={{}}]
		}
	}
}

\title{ESPnet-Codec: Comprehensive Training and Evaluation of Neural Codecs for Audio, Music, and Speech}
\name{\begin{tabular}[c]{@{}c@{}c@{}c@{}c@{}c@{}}
Jiatong Shi$^{1*}$, Jinchuan Tian$^{1*}$, Yihan Wu$^{1,2*}$, Jee-weon Jung$^{1\dagger}$, Jia Qi Yip$^{3\dagger}$, Yoshiki Masuyama$^{4\dagger}$, \\ William Chen$^{1\dagger}$, Yuning Wu$^{2\dagger}$, Yuxun Tang$^{2\dagger}$, Massa Baali$^{1}$, Dareen Alharthi$^{1}$, Dong Zhang$^{6}$,  Ruifan Deng$^{6}$, \\ Tejes Srivastava$^{7}$, Haibin Wu$^{7}$, Alexander H. Liu$^{9}$, Bhiksha Raj$^{1}$, Qin Jin$^{2}$, Ruihua Song$^{2}$, Shinji Watanabe$^{1}$
\end{tabular}
\thanks{$^*$co-first authors. \ $^{\dagger}$co-second authors. }
}
\address{\begin{tabular}[c]{@{}c@{}c@{}}
$^{1}$ Carnegie Mellon University, $^{2}$ Renmin University of China,  \\
$^{3}$ Nanyang Technological University, $^{4}$ Tokyo Metropolitan University, $^{5}$ Fudan University, \\
$^{6}$ University of Chicago, $^{7}$ National Taiwan University, $^{8}$ Massachusetts Institute of Technology
\end{tabular}}

\begin{document}
\ninept
\maketitle

\begin{abstract}
Neural codecs have become crucial to recent speech and audio generation research. In addition to signal compression capabilities, discrete codecs have also been found to enhance downstream training efficiency and compatibility with autoregressive language models. However, as extensive downstream applications are investigated, challenges have arisen in ensuring fair comparisons across diverse applications. To address these issues, we present a new open-source platform ESPnet-Codec, which is built on ESPnet and focuses on neural codec training and evaluation. ESPnet-Codec offers various recipes in audio, music, and speech for training and evaluation using several widely adopted codec models. Together with ESPnet-Codec, we present VERSA, a standalone evaluation toolkit, which provides a comprehensive evaluation of codec performance over 20 audio evaluation metrics. Notably, we demonstrate that ESPnet-Codec can be integrated into six ESPnet tasks, supporting diverse applications.
\end{abstract}


\section{Introduction} \label{sec:intro}

Speech representation, derived from speech signals, is fundamental for various speech tasks, including automatic speech recognition (ASR) and text-to-speech (TTS). 
Recently, self-supervised learning~(SSL) has emerged as a promising method in speech representation learning, leveraging unlabeled data to learn useful representations that surpass previous state-of-the-art results on various benchmarks~\cite{hsu2021hubert, chen2022wavlm, chiu2022self, shi2024multiresolution, superb, mlsuperb}.
However, continuous SSL representations in downstream tasks often have scalability issues related to storage and computation due to their higher dimensionalities (e.g., $1,024$ of WavLM-Large) compared to conventional acoustic features (e.g., $48$ of mel spectrogram)~\cite{chang2022distilhubert, chang23b_interspeech, chang2023exploring, shi2023bridging}. 
This has inspired a growing interest in learning discrete speech representation that aims to offer more efficient and compact representations for downstream tasks~\cite{hayashi2020discretalk, lee2022direct, shi2023enhancing, yan-etal-2023-espnet, yang2023towards, chang2023exploring, wang2023neural, copet2023simple, zhang2024speechtokenizer, yang2023uniaudio, zhang2023speechgpt, chang2024interspeech}.

One of the main streams in discrete speech representation is the neural codec type of approach. Initially proposed for audio compression, neural codecs are usually based on an encoder-decoder architecture, equipped with a quantizer that transmits intermediate representations to discrete codes~\cite{zeghidour2021soundstream, defossez2022high, kumar2024high}. While the original focus of neural codecs was signal compression, the interest in applying discrete representations for speech-related tasks has extended their usage into more downstream speech and audio tasks~\cite{wang2023neural, copet2023simple, zhang2024speechtokenizer, yang2023uniaudio, zhang2023speechgpt}.

Although neural codecs have become the basis for many downstream applications, several difficulties remain in comparing neural codecs in a controlled setup. These difficulties include, but are not limited to, variations in training data, training environment, and evaluation settings. Such challenges further complicate the evaluation of downstream applications, potentially impacting the corresponding scientific findings. To ensure a reasonable analysis of experimental findings, a framework that integrates various downstream tasks would be highly beneficial for future studies in both speech discrete representations and their downstream applications.

In this work, we propose a new toolkit, ESPnet-Codec, built on the ESPnet platform~\cite{watanabe2018espnet}. ESPnet-Codec currently supports five neural codec algorithms and will be further expanded.
The toolkit integrates seamlessly with existing ESPnet downstream tasks. Following the design principles of previous ESPnet toolkits~\cite{watanabe2018espnet, hayashi2020espnet, hayashi2021espnet2, shi2022muskits, jung2024espnet, lu2022espnet, li2021espnet, chen2023reducing}, ESPnet-Codec includes a core PyTorch-based library and a collection of recipes in audio, music, and speech. These recipes follow an all-in-one design, encompassing data preparation, model training, model inference, and evaluation.\footnote{All codebase, data preprocessing scripts, and experimental configurations are released at \href{https://github.com/espnet/espnet}{ESPnet}. Pre-trained models are released at \href{https://huggingface.co/models?other=codec&sort=trending&search=espnet}{Huggingface}.}

In addition to releasing ESPnet-Codec, this paper also provides an extensive comparative study of neural codecs and their applications. We propose another standalone toolkit named VERSA~(Versatile Speech and Audio Evaluation toolkit), which is able to evaluate a wide array of existing codec algorithms for speech/audio/music generation tasks. 
With VERSA, we offer a comprehensive analysis within a highly controlled experimental environment. While demonstrating that ESPnet-Codec has comparable performance to existing toolkits, our experiments also show that no single model can achieve the best performance across all metrics, suggesting the necessity for comprehensive evaluation toolkits like VERSA. Furthermore, a major benefit of ESPnet-Codec is its tight integration with various downstream tasks. We test pre-trained codecs across six downstream tasks and demonstrate how codec performance correlates with effectiveness in these tasks. While previous codec-based applications have mostly focused on generation tasks, we demonstrate feasible integration with understanding tasks and highlight the limitations of existing codec models.

\begin{table*}[t]
    \centering
        \caption{A comparison of existing codec-related toolkits. Codec types refer to the framework of different codec models, while recipes indicate whether the toolkit provides dataset-related training, inference, and evaluation pipelines. (The information is collected in June, 2024.)}
        \vspace{-10pt}
    \resizebox{\textwidth}{!}{
    \begin{tabular}{ l | c | c | c | c | c | c |c | c | c | c| c |  c | c | c| c |c  }
        \toprule
        \multirow{2}{*}{\textbf{Toolkit}} & \multirow{2}{*}{\textbf{Release}} &  \multirow{2}{*}{\textbf{\# Codecs}} & \multicolumn{5}{c|}{\textbf{Open-source Level}}  & \multirow{2}{*}{\textbf{\# Recipes}} & \multicolumn{8}{c}{\textbf{Supported Downstream Tasks using Codec}}  \\ 
        \cmidrule{4-8}\cmidrule{10-17}
        & & & Data & Training & Inference & Evaluation & Weights &  & ASR & TTS & TTM & TTA & SSE & SVS & SPK & SSL\\
        \midrule
        Encodec~\cite{defossez2022high} & 2022/10 & 1 & \xmark & \xmark & \cmark & \xmark & \cmark & 0 & \xmark & \xmark & \xmark & \xmark & \xmark & \xmark & \xmark & \xmark  \\
        \midrule
        AudioDec~\cite{wu2023audiodec} & 2023/05 & 1  & \cmark & \cmark & \cmark & \xmark & \cmark  & 0 & \xmark & \xmark & \xmark & \xmark & \cmark & \xmark & \xmark & \xmark \\ 
        \midrule
        AcademiCodec~\cite{yang2023hifi} & 2023/05 & 3 &  \xmark & \cmark & \cmark & \cmark & \cmark & 0 & \xmark & \xmark & \xmark & \xmark & \xmark & \xmark & \xmark & \xmark \\ 
        \midrule
        DAC~\cite{kumar2024high} & 2023/06 & 2 & \xmark & \cmark & \cmark & \cmark & \cmark &  0  & \xmark & \xmark & \xmark & \xmark & \xmark & \xmark & \xmark & \xmark \\ 
        \midrule
        AudioCraft~\cite{copet2023simple} & 2023/06 & 1 & \cmark & \xmark & \cmark & \cmark & \xmark & 0  & \xmark & \xmark & \cmark & \cmark & \xmark & \xmark & \xmark & \xmark \\ 
        \midrule
        SpeechTokenizer~\cite{zhang2024speechtokenizer} & 2023/08 & 1 & \xmark & \cmark & \cmark & \xmark & \cmark & 0  & \xmark & \cmark & \xmark & \xmark & \xmark & \xmark & \xmark & \xmark \\ 
        \midrule
        FunCodec~\cite{du2024funcodec} & 2023/09 &  3 & \cmark & \cmark & \cmark & \cmark & \cmark & 1  & \xmark & \cmark & \xmark & \cmark & \xmark & \xmark & \xmark & \xmark  \\ 
        \midrule
        Amphion~\cite{zhang2023amphion} & 2023/12 & 1 & \xmark & \xmark & \cmark & \xmark & \cmark & 0 & \xmark & \cmark & \xmark & \xmark & \xmark & \xmark & \xmark & \xmark \\ 
        \midrule
        \midrule
        ESPnet-Codec & 2024/06 & 5 & \cmark & \cmark & \cmark & \cmark & \cmark  & 7 & \cmark & \cmark & \xmark & \xmark & \cmark & \cmark & \cmark & \cmark \\ 
        \bottomrule
    \end{tabular}
    }
    \label{tab:codec_toolkits}
        \vspace{-15pt}
\end{table*}

\section{Related Works} \label{sec:reated work}

While existing open-source works in neural codecs have made significant contributions to the community~\cite{defossez2022high, kumar2024high, du2024funcodec, zhang2023amphion, copet2023simple, zhang2024speechtokenizer, wu2023audiodec, yang2023hifi}, they still face issues in codec training and integration with other tasks, as detailed in Table~\ref{tab:codec_toolkits}. First, although most toolkits provide codec training scripts, they usually do not provide all-in-one recipes with data preparation, training, inference, evaluation, and pre-trained model weights available. 
The limitation then complicates the comparison of models from different toolkits~\cite{defossez2022high, copet2023simple, zhang2023amphion, yang2023hifi, kumar2024high, zhang2024speechtokenizer}. Notably, some of the best publicly available pre-trained models are not associated with public datasets~\cite{defossez2022high, du2024funcodec}. 

Moreover, there has been a need for a unified yet comprehensive evaluation toolkit for signal quality measures. Existing toolkits either do not support in-toolkit evaluation or provide limited evaluation metrics. In Table~\ref{tab:codec_toolkits}, AcademiCodec, DAC, AudioCraft, and FunCodec provide evaluation protocols, but all of them support only a few metrics for evaluation.\footnote{AcademiCodec: PESQ~\cite{rix2001perceptual} and STOI~\cite{taal2011algorithm}; DAC: training losses, SI-SDR~\cite{le2019sdr}, and VISQOL~\cite{hines2015visqol}; AudioCraft: VISQOL and SI-SNR~\cite{le2019sdr}; FunCodec: VISQOL and word error rates from ASR systems.} When using only one or two metrics, potential biases can be easily introduced (see Sec.~\ref{sec:exp} for analysis). In contrast, using a wider range of metrics allows for a more comprehensive analysis and understanding of different codec models.

Additionally, current frameworks' integrations of downstream tasks only focus on limited generation tasks. However, considering the growing needs in spoken language modeling and the diverse usage of codec tokens~\cite{lee2022direct, shi2023enhancing, yan-etal-2023-espnet, yang2023towards, chang2023exploring, zhang2023speechgpt}, incorporating diverse tasks can benefit not only direct applications but also our understanding of different codec algorithms in views from downstream applications (see Sec.~\ref{sec:exp} for some analysis).

To address these limitations, we propose ESPnet-Codec, built on the foundation of existing ESPnet tasks, including ASR, TTS, speaker recognition (SPK), speech separation\&enhancement~(SSE), singing voice synthesis (SVS), and self-supervised learning pre-training~(SSL)~\cite{watanabe2018espnet, hayashi2020espnet, hayashi2021espnet2, shi2022muskits, jung2024espnet, lu2022espnet, li2021espnet, chen2023reducing}. With ESPnet-style recipes, this toolkit provides transparent comparisons among codec algorithms by re-implemention within the ESPnet ecosystem. The codec models can be thoroughly evaluated using the latest speech evaluation framework by VERSA. Additionally, the model supports integration with six existing ESPnet tasks through a discretization interface. 

\section{Functionalities} \label{sec:function}


\subsection{General Neural Codec Framework}

ESPnet-Codec follows general neural codec frameworks: The source input for the codec model is a sampled audio signal $S \in \mathbb{R}^{1 \times T_s}$ with a length of $T_s$ samples.\footnote{Depending on the codec model, preprocessing such as transformations to the frequency domain can be conducted for following codec modules.} The codec model $M$ has an $\mathrm{Encoder}(\cdot)$, a quantizer $\mathrm{Quantizer}(\cdot)$, and a decoder $\mathrm{Decoder}(\cdot)$. In the quantizer, we define a number of $L$ codebooks (i.e., \{$\mathcal{B}_1, ..., \mathcal{B}_L$\}) where the $i^{\text{th}}$ codebook $\mathcal{B}_i = \{1, 2, ..., B_i\}$ has $B_i$ codes. The forward process of codec $M$ is a typical serial pipeline.
The encoder first converts $S$ into hidden states $E \in \mathbb{R}^{D \times T_e}$ with a sequence length of $T_e$ and a dimension of $D$ (i.e., $E = \mathrm{Encoder}(S)$). Based on $E$, the quantizer generates discrete codes $C \in (\mathcal{B}_1, \mathcal{B}_2, ..., \mathcal{B}_L)^{T_c}$ with $T_c$ discretized frames, where $C$ can be interpreted into $\hat{E} \in \mathbb{R}^{D \times T_e}$ with the codebooks \{$\mathcal{B}_1, ..., \mathcal{B}_L$\} (i.e., $(C, \hat{E})  = \mathrm{Quantizer}(E | \mathcal{B}_1, \mathcal{B}_2, ..., \mathcal{B}_L)$). Lastly, the decoder utilizes $\hat{E}$ to obtain $\hat{S}$, aiming to reconstruct the original signal $S$ (i.e., $\hat{S} = \mathrm{Decoder}(\hat{E})$).

\noindent \textbf{Sound Representation}. ESPnet-Codec supports the use of both raw waveform and spectrogram from short-time Fourier transform~(STFT). The spectral option provides flexibility in supporting codecs based on spectral properties or complex domain~\cite{liu2023high, du2024funcodec}.

\noindent \textbf{Encoder and Decoder Architecture}. Due to the efficiency concern, existing codec algorithms mostly use convolutional encoders and decoders for pre/post quantization processing. Follow existing codec implementations~\cite{copet2023simple, defossez2022high}, ESPnet-Codec supports SEANet encoder and decoder~\cite{li2021real} and their variants in \cite{wu2023audiodec, yang2023hifi, kumar2024high, zhang2024speechtokenizer}. For additional decoder types, as suggested in \cite{wu2023audiodec}, ESPnet-Codec also supports non-symmetric decoders following the vocoder designs, sourced from ESPnet2-TTS~\cite{hayashi2021espnet2}.

\noindent \textbf{Quantization Algorithm}. ESPnet-Codec primarily supports residual vector quantization (RVQ) as the quantization algorithm, where each VQ follows the VQ-variational auto-encoder (VQ-VAE)-based optimization with exponential moving average~\cite{razavi2019generating}. A few variants of RVQ, including Group-RVQ (GRVQ) proposed in~\cite{yang2023hifi}, are also supported.

\noindent \textbf{Training and Loss Functions}. We primarily select the generative adversarial network (GAN)-based training framework for the codec trainer. Aligned with existing toolkits~\cite{du2024funcodec, copet2023simple}, ESPnet-Codec has training objectives in three focuses: reconstruction, adversarial, and quantization focuses. 

For reconstruction losses, ESPnet-Codec offers time-domain speech signal loss and frequency-domain mel spectrogram losses:
\vspace{-5pt}
\begin{equation}
    \mathcal{L}_{\text{rec.}}(S, \hat{S}) = ||S - \hat{S}||_{\text{norm}} + \frac{1}{|\mathcal{A}|}\sum_{a \in \mathcal{A}} ||\mathrm{M}_{a}(S) - \mathrm{M}_{a}(\hat{S})||_{\text{norm}},
    \vspace{-5pt}
\end{equation}
where $||\cdot||_{\text{norm}}$ can be either L1 norm, L2 norm, or their combination; $\mathcal{A}$ is a set of scales for different mel spectrogram;\footnote{The scale here refers to the STFT parameters. Default in our models, we use window sizes of \{$2^5, 2^6, ..., 2^11$\} with the frame shifts of $\frac{1}{4}$ window sizes.} $\mathrm{M}_a$ is the corresponding mel spectrogram extractor to the scale $a$. Additionally, we support distillation losses from teacher-student learning, which utilize pre-trained SSL models as in \cite{zhang2024speechtokenizer}.

For adversarial losses, the generator has both the discriminator losses and a feature matching loss:
\vspace{-5pt}
\begin{equation}
\label{eq: adv_loss}
    \mathcal{L}_{\text{gen.}} = \frac{1}{K}\sum_k\max(0, 1 - \mathrm{D}_k(\hat{S})) + \frac{1}{K \cdot R}\sum_{k,r}||\mathrm{D}_k^r(S) - \mathrm{D}_k^r(\hat{S})||_1,
\end{equation}
where $\mathrm{D}_k$ is the $k^{\text{th}}$ discriminator, $r$ represents the $r^{\text{th}}$ layer, $K$ is the total number of discriminator, and $R$ is the number of outputs in the discriminator. Discriminators are optimized with:
\vspace{-5pt}
\begin{equation}
    \mathcal{L}_{\text{disc.}} = \frac{1}{K}\sum_k[\max(0, 1 + \mathrm{D}_k(\hat{S})) + \max(0, 1 - \mathrm{D}_k(S))].
        \vspace{-5pt}
\end{equation}
ESPnet-Codec supports up to six variants of discriminators. Inherited from ESPnet2-TTS~\cite{hayashi2021espnet2}, we support the multi-resolution STFT discriminator from ParallelWaveGAN~\cite{yamamoto2020parallel}, the filter-bank random window discriminator (FB-RWD) from StyleMelGAN~\cite{mustafa2021stylemelgan}, the multi-scale discriminator (MSD) and the multi-period discriminator (MPD) from HiFi-GAN~\cite{kong2020hifi}, the simple STFT-based discriminator from SoundStream~\cite{zeghidour2021soundstream}, the multi-scale complex STFT discriminator from Encodec~\cite{defossez2022high}, and the multi-scale multi-period multi-band~(MSMPMB) discriminator from DAC~\cite{kumar2024high}.

For quantization losses, we consider commitment losses from both the whole quantizer and different levels of the quantizer as in:
\vspace{-10pt}
\begin{equation}
    \mathcal{L}_{\text{quan.}} = ||E - \hat{(E)}||_1 + \frac{1}{L}\sum_{i=1}^{L} ||Q_{i-1} - \mathrm{VQ}_i(Q_{i-1})||_\text{norm},
    \vspace{-10pt}
\end{equation}
where $Q_i$ is the $i^{\text{th}}$ hidden states to be encoded and $Q_0$ is identical to $E$. $L$, as defined above, is the number of codebooks (i.e., the level of the RVQ quantizer). $\mathrm{VQ}_i$ is the $i^{\text{th}}$ quantizer.


\subsection{Example Neural Codec Models}
\label{ssec: example models}

\noindent \textbf{SoundStream}: SoundStream~\cite{zeghidour2021soundstream} represents the initial effort in applying RVQ to neural codec learning. While the generator has a SEANet-based encoder-decoder and the RVQ quantizer, the model has two discriminators consisting of a waveform-based discriminator and a spectral discriminator. The losses in SoundStream include adversarial losses and feature-matching losses from the discriminators, as well as a multi-resolution mel loss.

\noindent \textbf{Encodec}: Encodec~\cite{defossez2022high} follows a similar architecture to SoundStream but uses a multi-scale STFT discriminator to replace the STFT discriminator. During training, the discriminator is not updated at $p_{\text{skip}}$ probability. Furthermore, a loss balancer is used to automatically synchronize the loss scales.

\noindent \textbf{DAC}: DAC~\cite{kumar2024high} extends Encodec with a few updates, including the utilization of the snake activation function, improved RVQ with factorized codes and L2-normalized codes, increased quantizer dropout, and the use of an MSMPMB discriminator.

\noindent \textbf{FunCodec}: While FunCodec is proposed as a toolkit~\cite{du2024funcodec}, it introduces the interface of using complex STFT to train codec in the frequency domain. The codec model named FunCodec in ESPnet follows the ``FreqCodec'' configuration in the FunCodec repository, which utilizes a multi-scale STFT discriminator.

\noindent \textbf{HiFi-Codec}: HiFi-Codec~\cite{yang2023hifi} improves on Encodec by using GRVQ to split the hidden states into multiple sub-groups, where each sub-group applies RVQ for reconstruction independently. Moreover, HiFi-Codec employs three discriminators, including MSD, MPD, and multi-scale STFT discriminators.


\subsection{Speech and Audio Quality Evaluation with VERSA}
\label{ssec: versa}

As discussed in Sec.~\ref{sec:reated work}, previous codec-related toolkits typically support a limited set of evaluation metrics, leading to discrepancies in metric selection. Moreover, reported numbers can vary due to differences in hyper-parameter setups or the downstream models used for evaluation. Additionally, some metrics, while demonstrating superior quality in the contexts for which they were designed, may not be as robust under diverse conditions (see Section~\ref{sec:exp} for details). Relying on a limited set of metrics is risky for codec model selection, especially when codec models often serve as fundamental modules for other systems. To address these issues, we introduce a standalone Versatile Speech and Audio Evaluation toolkit (VERSA) alongside ESPnet-Codec to unify the evaluation framework. VERSA supports both intrusive (full-reference), non-intrusive (no-reference) metrics, and other perceptual metrics that are not directly related to signal-level information. The intrusive category includes both non-learning and learning-based measures, while the non-intrusive category and other perceptual metrics consist solely of learning-based measures.

For intrusive metrics, we support non-learning measures, including mel cepstral distortion (MCD), F0 root mean square error~(F0-RMSE), F0 Pearson correlation (F0-CORR), scale-invariant signal-to-noise ratio (SI-SNR)~\cite{le2019sdr}, convolutive transfer function invariant signal-to-distortion ratio (CI-SDR)~\cite{boeddeker2021convolutive}, perceptual evaluation of speech quality~(PESQ)~\cite{rix2001perceptual}, and short-time objective intelligibility~(STOI)~\cite{taal2011algorithm}. For learning-based metrics, VERSA supports virtual speech quality objective listener~(VISQOL)~\cite{hines2015visqol}, speech BLEU score~(D-BLEU)~\cite{saeki2024speechbertscore}, discrete speech distance~(D-Distance)~\cite{saeki2024speechbertscore}, and speech BERT score~(S-BERT)~\cite{saeki2024speechbertscore}. For non-intrusive metrics, we support deep noise suppression mean opinion score (DNSMOS)~\cite{reddy2021dnsmos}, UTokyo-SaruLab system for VoiceMOS Challenge~(UTMOS)~\cite{saeki2022utmos}, packet loss concealment MOS~(PLCMOS)~\cite{dienerplcmos}, and singing MOS~(SingMOS)~\cite{tang2024singmos}. For other perceptual metrics, VERSA includes character/word error rate (CER/WER) by pre-trained speech recognition models from either ESPnet or OpenAI-Whisper~\cite{watanabe2018espnet, radford2023robust}, Speaker similarity~(SPK-SIM) powered by more than ten speaker-embedding extractors based on ESPnet-SPK~\cite{jung2024espnet}. By providing this comprehensive suite of evaluation tools, VERSA aims to standardize and streamline the assessment of codec performance across different models and applications.

\begin{table}[t]
\centering
\caption{Evaluation metrics supported in the VERSA toolkit. Evaluation metrics are listed in abbreviations. Details are in Sec.~\ref{ssec: versa}. }
\vspace{-10pt}
\resizebox{\linewidth}{!}{
\begin{tabular}{l|c|c}
\toprule
\textbf{Dependency} & \textbf{Base} & \textbf{Evaluation Metrics} \\ 
\midrule
\multirow{3}{*}{Intrusive} & \multirow{2}{*}{Non-Learning} & MCD, F0-RMSE, F0-CORR, SI-SNR,   \\ 
&  & CI-SDR, PESQ, STOI \\ 
\cmidrule{2-3}
& \multirow{3}{*}{Learning} & VISQOL, D-BLEU, D-Distance, S-BERT  \\ 
 \cmidrule{1-1} \cmidrule{3-3}
\multirow{1}{*}{Non-intrusive}   &  &  DNSMOS, UTMOS, PLCMOS, SingMOS \\ 
 \cmidrule{1-1} \cmidrule{3-3}
 Other Perceptual &  & CER/WER, SPK-SIM \\
\bottomrule
\end{tabular}
}
\label{tab:framework}
\vspace{-15pt}
\end{table}

\subsection{Downstream Application}

This section showcases several downstream applications of pre-trained ESPnet-Codec models. While we demonstrate the ease and practicality of applying the codec to various downstream tasks, these tasks also assess the generalizability, robustness, and efficiency of different speech neural codecs.

\noindent \textbf{ASR} transcribes text from speech. Recent studies show that ASR based on discrete representation (i.e., discrete ASR) can achieve efficient training, reduced storage constraints, and improved performance over spectral features~\cite{chang2023exploring, chang23b_interspeech, wang2023viola, chen2024loss, yang2023towards, chang2024interspeech, ye2023asq, shi2024mmm}. While most previous works focus on discrete units from pre-trained SSL models, existing discrete ASR systems using neural codecs are typically presented in a multi-task manner with large-scale training data~\cite{wang2023viola, gupta2024exploring}.

Supported by ESPnet~\cite{chang2023exploring}, the framework for discrete ASR, pre-trained Codec models can be easily integrated into the discrete ASR task as model input. The discrete tokens from neural codecs can either be mapped to embeddings or looked up from the existing codebooks in the codec models. Embeddings from multi-level codebooks are then fed to the downstream ASR model for ASR training, enjoying the full capacity of ESPnet ASR architectures.

\noindent \textbf{TTS}  has been a major application for neural codecs. Before the utilization of discrete representation, TTS was dominated by non-autoregressive (NAR) algorithms~\cite{ren2020fastspeech, kim2021conditional} due to their efficiency, controllability, and stable performance over their autoregressive~(AR) counterparts~\cite{shen2018natural}. However, the introduction of discrete neural codecs has brought additional attention to AR modeling~\cite{wang2023speechx, wang2023neural, yang2023uniaudio, anastassiou2024seed}, which has proven versatile in generating expressive speech with robustness in zero-shot scenarios. Meanwhile, recent works have shown that NAR TTS can also benefit from using pre-trained codecs in their modeling~\cite{junaturalspeech, yang2024simplespeech}. In this work, on top of foundations from ESPnet-TTS~\cite{hayashi2020espnet, hayashi2021espnet2} we develop two discrete TTS tasks with ESPnet-Codec to support both NAR and AR TTS.

\noindent \underline{\textit{NAR TTS}}. For NAR TTS, we follow the discrete TTS approach proposed in~\cite{shi2024mmm}. The model employs a Fastspeech2-like backbone as in~\cite{chang2024interspeech} with a variational auto-encoder-based approach to achieve NAR TTS without duration supervision. The prediction targets for the NAR TTS are codecs from reference speech signals. Multi-speaker TTS is achieved by using pre-trained speaker embeddings.

\noindent \underline{\textit{AR TTS (Speech LM)}}. For AR TTS, we implement VALL-E~\cite{wang2023neural}. The model uses an AR decoder-only model to predict the first stream of codecs from the given text and a target speaker speech prompt. Then, the following streams are predicted in an NAR fashion based on the first codec stream.

\noindent \textbf{SPK}  (i.e., the SPK task in the paper) focuses on identifying the speaker from speech signals. While existing SPK models primarily depend on spectral features, raw waveform, or SSL features~\cite{jung2022pushing, desplanques2020ecapa, chen2022wavlm}, recent work has shown the potential of discrete neural codecs as an alternative to existing speech features~\cite{puvvada2024discrete}.

As an important attribute of speech signals, ESPnet-Codec can also be adopted for the SPK task by integrating with ESPnet-SPK~\cite{jung2024espnet}, simply by replacing the acoustic input with codecs.

\noindent \textbf{SSE} aims at extracting the target speech from a mixture of overlapping speakers with background noise.
Typical approaches to SSE have leveraged STFT~\cite{Wang2022TFGridNetIF,Richter2023}, a trainable encoder~\cite{Zhao2023MossFormer2CT, Yip2023SPGMPL}, and a pre-trained SSL~\cite{Song2022slt} to obtain redundant representation.
Although a recent work~\cite{yip2024towards} has shown the potential SSE on neural codecs, compressed representations tend to cause degraded performance on objective metrics~\cite{Subakan2022ExploringSM}.
Hence, a renewed emphasis on perceptual evaluation is required to clarify the advantages of codecs.

ESPnet-Codec supports SSE on top of ESPnet-SE~\cite{yip2024towards} and implements additional flexibility in switching between activation functions, masking, and mapping approach, which have been identified as factors impacting the performance of codec-based SSE.

\noindent \textbf{SVS} targets the synthesis of singing voices from musical inputs. Compared to the similar task of TTS, SVS has strict requirements for rhythm control and a greater sensitivity to high-frequency information in the signal. The concept of discrete SVS was introduced in the discrete speech challenge at Interspeech 2024~\cite{chang2024interspeech}, prompting several investigations into the application of pre-trained SSL tokens to SVS~\cite{wu2024toksing, tang2024singomd}. The winning system, ``TokSing," demonstrated superior performance and efficiency compared to models using spectral features or VAE-based latent features~\cite{wu2024toksing, zhang23e_interspeech, shi2021sequence}.

Building on Muskits~\cite{shi2022muskits}, we implement the TokSing architecture within the context of neural codecs using ESPnet-Codec. Similar to TokSing, we predict multi-stream neural codecs in parallel, serving as the training targets for the SVS task.

\noindent \textbf{SSL} in speech/audio signals has proven to be an effective way of utilizing unlabeled data, especially in extracting semantic features~\cite{hsu2021hubert, chen2022wavlm, chiu2022self, chen2023reducing}. Existing speech/audio SSL models mostly rely on the raw waveform, which can incur a large computational overhead in the feature extractor~\cite{lin2023melhubert, parcollet2023efficiency, lugo2024towards, yang2023fast}.

While existing methods focus on using spectral features to replace the convolutional feature extractor~\cite{lin2023melhubert, parcollet2023efficiency, lugo2024towards, yang2023fast}, ESPnet-Codec connects to the ESPnet-SSL task~\cite{chen2023reducing} to support Codec-based SSL for the first time, replacing the raw waveform with discrete codes.
\section{Experiments} \label{sec:exp}
\begin{table*}[ht]
\centering
\caption{LibriTTS codec performance comparison at \{16 / 24\}kHz. * indicates models trained using existing open-source codebases.}
        \vspace{-10pt}
\resizebox{\textwidth}{!}{
\begin{tabular}{l|c|c|c|c|c|c|c|c|c|c|c|c|c|c}
\toprule
\multirow{3}{*}{\textbf{Model}} & \multicolumn{9}{|c|}{\textbf{Intrusive}} & \multicolumn{3}{|c|}{\textbf{Non-Intrusive}} & \multicolumn{2}{|c}{\textbf{Perceptual}}  \\
\cmidrule{2-15}
 & \multicolumn{5}{|c|}{\textbf{Non-learning}} & \multicolumn{9}{|c}{\textbf{Learning}}    \\
\cmidrule{2-15}
 & \textbf{MCD $\downarrow$} & \textbf{F0-RMSE $\downarrow$} & \textbf{F0-CORR $\uparrow$} & \textbf{STOI $\uparrow$} & \textbf{PESQ $\uparrow$}  & \textbf{D-BLEU $\uparrow$} & \textbf{D-Distance $\uparrow$} & \textbf{S-BERT $\uparrow$} &  \textbf{VISQOL $\uparrow$} & \textbf{UTMOS $\uparrow$} & \textbf{PLCMOS $\uparrow$} & \textbf{DNSMOS $\uparrow$} & \textbf{WER $\downarrow$} & \textbf{SPK-SIM $\uparrow$} \\
\midrule
DAC* \cite{kumar2024high} & 13.85 / \textbf{3.17} & 61.03 / 52.09 & \textbf{0.61} / \textbf{0.72} & \textbf{0.98} / \textbf{0.99} & \textbf{3.81} / 3.45  & \textbf{0.88} / \textbf{0.93} & \textbf{0.91} / \textbf{0.94} & \textbf{0.98} / \textbf{0.99} & 2.90 / 4.60 & 3.81 / 3.95  & 4.01 / 4.19 & 3.05 / 3.02 & 2.9 / 2.7 & 0.85 / \textbf{0.97} \\ 
Encodec* \cite{defossez2022high} & \phantom{0}4.84 / 3.94 & \textbf{42.04} / \textbf{38.00} & 0.59 / 0.65 & 0.97 / 0.98 & 3.45 / \textbf{4.09}  & 0.85 / 0.89 & 0.90 / 0.92 & \textbf{0.98} / \textbf{0.99} & 4.49 / \textbf{4.73} & 3.86 / \textbf{4.08} & 4.16 / \textbf{4.27} &  3.06 / 3.09 & 2.4 / 2.4 & 0.86 / 0.95 \\
FunCodec* \cite{du2024funcodec} & \phantom{0}5.40 / 5.52  & 43.19 / 43.81  & 0.55 / 0.55  & 0.90 / 0.93 & 3.23 / 3.11  & 0.78 / 0.79 & 0.87 / 0.87 & 0.95 / 0.96 & 4.57 / 4.52 & 3.84 / 3.84 & \textbf{4.48} / 4.22 & \textbf{3.15} / \textbf{3.14} & 2.2 / \textbf{2.1} & \textbf{0.93} / 0.92 \\
SpeechTokenizer* \cite{zhang2024speechtokenizer} & \phantom{0}5.46 / \phantom{0}-\phantom{00} & 43.32 / \phantom{00}-\phantom{00} & 0.55 / \phantom{0}-\phantom{00} & 0.93 / \phantom{0}-\phantom{00} & 2.62 / \phantom{0}-\phantom{00}  & 0.80 / \phantom{0}-\phantom{00} & 0.87 / \phantom{0}-\phantom{00} & 0.95 / \phantom{0}-\phantom{00} & 4.36 / \phantom{0}-\phantom{00} & 3.80 / \phantom{0}-\phantom{00} & 4.08 / \phantom{0}-\phantom{00} & 3.06 / \phantom{0}-\phantom{00} & 2.2  /  \phantom{0}-\phantom{0} & 0.62 / \phantom{0}-\phantom{00}  \\
\midrule
SoundStream & \phantom{0}4.64 / 4.28 & 43.02 / 41.08 & 0.57 / 0.57 & 0.95 / 0.93  & 2.83 / 2.54   & 0.83 / 0.78 & 0.89 / 0.86 & 0.96 / 0.94 & 4.45 / 4.21 & 3.76 / 3.58 & 4.38 / 4.03 & 3.10 / 2.97 & \textbf{2.1} / \textbf{2.1}  & 0.84 / 0.76 \\
Encodec & \phantom{0}\textbf{3.73} / 4.26 & 43.57 / 46.12 & 0.55 / 0.53 & 0.97 / 0.97 & 3.37 / 3.25  & 0.87 / 0.87 & 0.91 / 0.91 & \textbf{0.98} / 0.97 & \textbf{4.65} / 4.55 & \textbf{3.92} / 3.93 & 4.45 / 4.17 & 3.11 / 3.12 & \textbf{2.1} / \textbf{2.1} & 0.90 / 0.86\\ 
DAC & \phantom{0}5.16 / 4.98 & 65.20 / 62.92 & 0.56 / 0.53 & 0.95 / 0.97 & 3.00 / 3.43 & 0.85 / 0.87 & 0.89 / 0.90 & 0.97 / 0.97 & 4.48 / 4.21 & 3.78 / 3.93 & 4.28 / 4.07 & 3.04 / 3.05 & 2.9 / 2.8 & 0.76 / 0.83\\
HiFi-Codec & \phantom{0}6.03 / 6.37 & 42.33 / 42.58 & 0.56 / 0.56 & 0.94 / 0.95 & 2.48 / 2.42 & 0.78 / 0.77 & 0.86 / 0.85 & 0.94 / 0.93 & 4.01 / 3.96 & 3.61 / 3.58 & 4.23 / 3.88 & 2.89 / 2.89 & 3.1 / 3.7 & 0.66 / 0.65  \\
Fun-Codec & \phantom{0}5.87 / 5.76 & 67.72 / 69.90 & 0.54 / 0.51 & 0.83 / 0.86 & 2.43 / 2.40 & 0.71 / 0.73 & 0.82 / 0.83 & 0.90 / 0.91 & 4.26 / 4.24 & 3.01 / 3.09 & 4.13 / 3.90 & 3.04 / 3.03 & 3.7 / 3.6 & 0.72 / 0.70 \\
\midrule
Ground Truth & - & - & - & - & - & - & - & - & - & 4.05 & 4.42 & 3.08  &  2.1 & - \\
\bottomrule
\end{tabular}}
\label{table:main_comparison_16khz}
\vspace{-10pt}
\end{table*}

\begin{table*}[ht]
\centering
\caption{AMUSE codec performance comparison on the \textit{\textbf{speech}} test set at \{16 / 44.1\}kHz.}
\vspace{-10pt}
\resizebox{0.95\textwidth}{!}{
\begin{tabular}{l|c|c|c|c|c|c|c|c|c|c|c|c|c}
\toprule
\multirow{3}{*}{\textbf{Model}} & \multicolumn{9}{|c|}{\textbf{Intrusive}} & \multicolumn{3}{|c|}{\textbf{Non-Intrusive}} & \multicolumn{1}{|c}{\textbf{Perceptual}}  \\
\cmidrule{2-14}
 & \multicolumn{5}{|c|}{\textbf{Non-learning}} & \multicolumn{8}{|c}{\textbf{Learning}}    \\
\cmidrule{2-14}
 & \textbf{MCD $\downarrow$} & \textbf{F0-RMSE $\downarrow$} & \textbf{F0-CORR $\uparrow$} & \textbf{STOI $\uparrow$} & \textbf{PESQ $\uparrow$} & \textbf{D-BLEU $\uparrow$} & \textbf{D-Distance $\uparrow$} & \textbf{S-BERT $\uparrow$} &  \textbf{VISQOL $\uparrow$} & \textbf{UTMOS $\uparrow$} & \textbf{PLCMOS $\uparrow$} & \textbf{DNSMOS $\uparrow$} & \textbf{SPK-SIM $\uparrow$} \\
\midrule
SoundStream & \textbf{4.77} / 5.87 & \textbf{42.95} / 46.67 & \textbf{0.51} / 0.49 & \textbf{0.92} / 0.94 & 2.71 / 2.79  & \textbf{0.71} / 0.73 & \textbf{0.88} / 0.88 & \textbf{0.95} / 0.96 & 4.51 / 4.49 & 1.95 / 1.93 & \textbf{3.78} / \textbf{3.65} & 2.66 / 2.59 & \textbf{0.87} / 0.82 \\
Encodec &  4.91 / 5.14 & 46.73 / 57.64 & 0.46 / 0.37 & 0.91 / 0.91 & 2.28 / 1.81  & 0.70 / 0.73 & 0.87 / 0.88 & 0.94 / 0.95 & \textbf{4.52} / 4.45 & 1.55 / 1.59 & 3.64 / 2.17 & 2.54 / 2.44 & 0.85 / 0.85  \\
DAC & 5.76 / \textbf{4.99} & 44.36 / \textbf{43.19} & \textbf{0.51} / \textbf{0.55} & 0.91 / \textbf{0.97} & \textbf{2.75} / \textbf{4.02} & 0.68 / \textbf{0.79} & 0.86 / \textbf{0.90} & 0.94 / \textbf{0.98} &  4.29 / \textbf{4.67} & \textbf{2.14} / \textbf{2.37} & 3.65 / 2.34 & \textbf{2.77} / \textbf{2.84} & 0.78 / \textbf{0.93}  \\
\midrule
Ground Truth & - & - & - & - & - & - & - & - & - & 2.45 & 3.85 & 2.79 & - \\
\bottomrule
\end{tabular}}
\label{table:main_comparison_large_speech}
\vspace{-10pt}
\end{table*}

\begin{table}[ht]
\centering
\caption{AMUSE codec performance comparison on the \textit{\textbf{audio}} test set at \{16 / 44.1\}kHz.}
\vspace{-10pt}
\resizebox{0.5\linewidth}{!}{
\begin{tabular}{l|c|c}
\toprule
\textbf{Model} & \textbf{MCD $\downarrow$} &  \textbf{VISQOL $\uparrow$}  \\
\midrule
SoundStream & \textbf{3.57} / 5.05  & \textbf{4.38} / 3.92 \\
Encodec & 3.59 / \textbf{4.90} & 4.32 / \textbf{4.03}\\
DAC & 3.59 / 4.97  & 4.23 / 3.90 \\
\bottomrule
\end{tabular}
}
\label{table:main_comparison_large_audio}
\end{table}

\begin{table}[ht]
\centering
\caption{AMUSE codec performance comparison on the \textit{\textbf{music}} test set at \{16 / 44.1\}kHz.}
\vspace{-10pt}
\resizebox{0.5\linewidth}{!}{
\begin{tabular}{l|c|c}
\toprule
\textbf{Model} & \textbf{MCD $\downarrow$}  & \textbf{VISQOL $\uparrow$} \\
\midrule
SoundStream & 3.86 / 5.60  & 3.77 / \textbf{3.93} \\
Encodec & 4.20 / 5.60  & \textbf{3.81} / 3.81 \\
DAC & \textbf{3.66} / \textbf{5.14}  & 3.52 / 3.67  \\
\bottomrule
\end{tabular}
}
\vspace{-15pt}
\label{table:main_comparison_large_music}
\end{table}

\begin{table}[ht]
\centering
\caption{CodecSUPERB benchmark performance (hidden set).+ indicate models trained on large-scale AMUSE dataset.}
\vspace{-10pt}
\resizebox{\linewidth}{!}{
\begin{tabular}{l|c|c|c|c|c|c|c|c}
\toprule
\multirow{3}{*}{\textbf{Model}} & \multicolumn{4}{|c|}{\textbf{Application}} & \multicolumn{4}{|c}{\textbf{Signal Metrics}}  \\
\cmidrule{2-9}
& \textbf{ASR} & \textbf{SPK} & \textbf{ER} & \textbf{AEC} & \multicolumn{3}{|c|}{\textbf{Speech}} & \textbf{Audio} \\
\cmidrule{2-9}
& WER $\downarrow$ & EER & ACC $\uparrow$ & ACC $\uparrow$ & PESQ $\uparrow$ & STOI $\uparrow$ & Mel Loss $\downarrow$ & Mel Loss $\downarrow$
\\
\midrule
SoundStream & 5.99 & 2.80 & 49.49 & 51.13 & 2.54 & 0.91 & 0.87 & 2.21 \\
Encodec & \textbf{5.58} & \textbf{2.20} & 52.53 & 61.58 & \textbf{2.99} & \textbf{0.95} & \textbf{0.74} & 1.02 \\
DAC &  6.33 & 3.40 & \textbf{55.56} & 52.65 & 2.51 & 0.91 & 0.90 & 1.25 \\
\midrule
SoundStream+ & 5.88 & 2.40 & 53.54 & \textbf{68.11} & 2.95 & 0.93 & \textbf{0.74} & \textbf{0.96}  \\
Encodec+ & 5.98 & 2.40 & 50.51 & 66.50 & 2.48 & 0.92 & 0.80 & 0.97 \\
DAC+ & 6.08 & 3.20 & 54.55 & 61.89 & 2.94 & 0.93 & 0.86 & 1.08  \\
\midrule
Original Audio & 5.28 & 1.60 & 59.60 & 78.01 & - & - & - & - \\
\bottomrule
\end{tabular}
}
\vspace{-15pt}
\label{tab:codecsuperb}
\end{table}

\subsection{Codec Dataset}

\textbf{LibriTTS (base dataset)}: Aligned with prior studies, we exemplify the toolkit on the LibriTTS dataset~\cite{zen2019libritts}. Specifically, we use the 460-hour training set for codec training and the test-clean set for evaluation. To study the effect of different sampling rates, we evaluate the model at both 16kHz and 24kHz sampling rates.

\noindent \textbf{AMUSE (large-scale dataset)}: Additionally, we include all the datasets used in DAC~\cite{kumar2024high} to gather a range of open-source high-quality datasets, forming a new fused dataset named the Audio, Music, and Speech Ensemble (AMUSE) corpus. All datasets in AMUSE have a sampling rate of 44.1kHz or higher. In addition to the datasets used in the DAC training set, AMUSE includes speech training data from AISHELL3~\cite{shi2021aishell}, Googlei18n-TTS corpora, and Mexican endangered languages~\cite{shi2021highland, shi2021leveraging, amith_audio_corpus_sierra, amith_totonac, amith_yoloxochitl_mixtec}.\footnote{As mentioned in \cite{kumar2024high}, its original collection of speech data has diverse qualities because of upsampling from low-sampling rate signals. Therefore, we intentionally add more TTS-focused data with known high quality into AMUSE to balance the training.} For music training data, AMUSE incorporates additional singing voice data from OpenSinger, StyleSing111, M4Singer, Kiritan-singing, Oniku Kurumi Utagoe database, Natsume Singing database, Opencpop, ACE-KiSing (excluding original voices), PJS, and JSUT singing~\cite{huang2021multi, dai2023singstyle111, zhang2022m4singer, ogawa2021tohoku, wang2022opencpop, shi2024singing, koguchi2020pjs, takamichi2020jsut}. We randomly sample 1,000 utterances from the training set to form the development set. For the test set, we collect the test sets of all datasets that explicitly include test sets. Given that the singing voice datasets are relatively small, we add a few more datasets to the test set, including PopCS, Ofuton P Utagoe database, and ACE-KiSing-original as additional test sets~\cite{liu2022diffsinger, shi2024singing}. We use the full music test set. Since the test set is relatively large, we subsample the test set to 3,000 utterances for audio and speech, respectively, using a fixed random seed. For AMUSE-related experiments, we investigate both 16kHz and 44.1kHz sampling rates.\footnote{A few recipes related to AMUSE will be released with ESPnet-Codec, which additionally provides codec options for \{audio, music, speech\} only scenarios for related usages. a Because of the space limitation, we do not present their results in the paper.}

\subsection{Codec Experimental Setup}

We conduct experiments with codecs in a controlled setting, including a base setting on LibriTTS and a large-scale setting on AMUSE. For the base setting, we train models using a single V100 GPU, a batch size of 8 with 1-second chunks, and 600k training steps. For the large-scale setting, models are trained using four Ampere GPUs, a batch size of 128 with 2-second chunks, and 600k training steps. Since all supported codecs use RVQ as their VQ strategy, we uniformly use 32-level codebooks, each with 1,024 codes. The training adopts multi-band learning with random bitrate control of {2, 4, 8, 16} kbps. The weights of losses for each model are tuned based on development sets. For codec configurations, we mostly follow the settings in their original papers or publicly released configurations\cite{zeghidour2021soundstream, defossez2022high, kumar2024high, du2024funcodec, yang2023hifi, zhang2024speechtokenizer}. A few specific changes are made based on hyperparameter tuning on the development sets.\footnote{For Encodec, we do not use the loss balancer as it did not show effectiveness on the development set. For DAC, we do not include factorization in the code space, as we empirically find the low-dimensional hidden states difficult to converge in our experiments.} For comparison, we also train the codec models on LibriTTS based on the original DAC, Encodec, FunCodec, and SpeechTokenizer~\cite{kumar2024high, copet2023simple, du2024funcodec, zhang2024speechtokenizer}.

For evaluation, resampling is conducted for metrics that are fixed to a specific sampling rate. The layer index and KMeans models for discrete-speech evaluation are based on recommended values from~\cite{saeki2024speechbertscore}. We use the speech version of VISQOL for speech evaluation, but the default version for audio and music evaluation~\cite{hines2015visqol}. The CER is calculated based on Whisper-large-V3 for codecs trained on LibriTTS, and the SPK-SIM is based on cosine similarity from pre-trained RawNet3~\cite{jung2022pushing} speaker embeddings from ESPnet-SPK~\cite{jung2024espnet}.

\subsection{Codec Results and Discussion}


The results of LibriTTS-base experiments are shown in Table~\ref{table:main_comparison_16khz}. In general, our reproduced codec models achieve performance comparable to existing codec implementations, with notably better performance on WER by Whisper. We also observe diverse differences across various evaluation metrics, as no single model achieves the best scores in all metrics. For instance, while the original DAC performs better in several metrics, it has much worse WER by Whisper compared to FunCodec. This finding supports our motivation for proposing VERSA, as discussed in Sec.~\ref{sec:reated work}. Within ESPnet-Codec models, Encodec shows the best performance in most metrics, while other models have their own strengths, such as SoundStream's superior F0 tracking and DAC's better PESQ in 24kHz speech.

Table~\ref{table:main_comparison_large_speech}, \ref{table:main_comparison_large_audio}, and \ref{table:main_comparison_large_music} show the evaluation results on AMUSE speech, audio, and music test sets, respectively. It is clear that large-scale data significantly impacts all three evaluated codec models, completely altering their relative performance on each metric. Notably, while Encodec performs well on LibriTTS, it does not generalize as well to large-scale data in speech (Table~\ref{table:main_comparison_large_speech}). Similar to the base experiments on LibriTTS, there is no agreement on all the evaluation metrics. In Table~\ref{table:main_comparison_large_speech}, Encodec has worse scores in three MOS scores but still has comparable VISQOL scores to other codec models. Similarly in Table~\ref{table:main_comparison_large_audio} and~\ref{table:main_comparison_large_music}, we observe a mixed result over different settings of the models.

\begin{table*}[t]
\centering
\caption{Comparison of codecs on ASR, TTS, SPK, SSE, SVS, and SSL (evaluated on ASR and SLU). + stands for codecs trained on AMUSE. Test-\{clean / other\} sets are reported for ASR. Models that failed to train are marked with \xmark.}
\vspace{-10pt}
\resizebox{\linewidth}{!}{
\begin{tabular}{l|c|c|c|c|c|c|c|c|c|c|c|c|c|c|c|c}
\toprule
\multirow{2}{*}{\textbf{Model}} & \textbf{ASR} & \multicolumn{3}{|c|}{\textbf{NAR-TTS}} & \multicolumn{3}{|c|}{\textbf{AR-TTS}} & \multicolumn{1}{|c|}{\textbf{SPK}} & \multicolumn{3}{|c|}{\textbf{SSE}} & \multicolumn{3}{|c|}{\textbf{SVS}} & \multicolumn{1}{|c|}{\textbf{SSL (ASR)}} & \multicolumn{1}{|c}{\textbf{SSL (SLU)}} \\
\cmidrule{2-17}
 & \textbf{WER $\downarrow$} & \textbf{WER $\downarrow$} & \textbf{UTMOS $\uparrow$} & \textbf{SPK-SIM $\uparrow$} & \textbf{WER  $\downarrow$} & \textbf{UTMOS $\uparrow$} & \textbf{SPK-SIM $\uparrow$} & \textbf{EER $\downarrow$} & \textbf{PESQ $\uparrow$} & \textbf{STOI $\uparrow$} & \textbf{DNSMOS $\uparrow$} & \textbf{MCD $\downarrow$} & \textbf{SACC $\uparrow$} & \textbf{SingMOS $\uparrow$} & \textbf{WER $\downarrow$} & \textbf{ACC $\uparrow$} \\
\midrule
SoundStream & 3.7 / 11.1 & \textbf{3.4} & 2.34 & 0.58 & \textbf{6.7} & 3.70 & 0.63 & 27.5 & 1.79 & 0.73 & 0.93 & - & - & - &  \xmark & \xmark  \\
Encodec & \textbf{3.6} / \textbf{10.0} & 4.3 & \textbf{2.35} & \textbf{0.59} & 7.7 & \textbf{3.85} & 0.63 & 15.7 & 1.85 & 0.76 & \textbf{0.95} & - & - & - & \xmark & \phantom{0}1.5\\
DAC & \textbf{3.6} / 10.3 & 5.2 & 2.12 & 0.54 & 10.2 & 3.75 & \textbf{0.66} & 29.5 & 1.73 & 0.74 & 0.87 & - & - & - &  \xmark & \xmark \\
\midrule
SoundStream+ & 3.7 / 10.2 & 4.7 & 1.84 & 0.57 & 9.8 & 2.97 & 0.61 & 15.9 &  1.76 & 0.73 & 0.81 & 8.82 & \textbf{0.60} & \textbf{2.92}  &  \xmark & \xmark \\
Encodec+ & 3.9 / 10.3 & 5.4 & 1.92 & 0.58& 8.6 & 2.32 & 0.62 & \textbf{14.1} & 1.24 & 0.51 & 0.62 & \textbf{8.51} & 0.58 & 2.86 & 28.1 & \textbf{68.9} \\
DAC+ & 4.1 / 10.9 & 6.2 & 1.82& 0.56 & 15.9 & 2.94 &0.64 & 24.6 & \textbf{2.00} & \textbf{0.78} & 0.87 & 9.26 & 0.49 & 2.58 & \xmark & \phantom{0}3.6 \\
\bottomrule
\end{tabular}
}

\label{table:codec_task_comparison}
\vspace{-15pt}
\end{table*}

\subsection{CodecSUPERB Evaluation}
While the above evaluation is conducted over VERSA-supported audio quality metrics, we also submit the pre-trained 16kHz codec results to the CodecSUPERB benchmark (hidden sets)~\cite{wu2024codec} for further evaluation on unknown hidden sets using a range of pre-trained models. In the application evaluation, resynthesized audio is input to pre-trained task-specific models for ASR, SPK, emotion recognition (ER), and audio event classification (AEC), with WER, equal error rate (EER), and accuracy (ACC) as evaluation metrics, respectively.

As presented in Table~\ref{tab:codecsuperb}, Encodec trained on base LibriTTS has achieved the best performance across most of the speech-related metrics, while SoundStream trained on AMUSE performs better on audio-related metrics. Although large-scale training with speech, audio, and music contributes to the performance of SoundStream and DAC, scaling up training does not necessarily improve the performance of Encodec, especially for speech-related metrics. This finding highlights the need for additional strategies when increasing data size and suggests that existing model architectures may exhibit severe sensitivity to diverse scenarios.



\subsection{Downstream Integration: ASR}

\noindent \textbf{Setups}: The discrete ASR experiments are conducted on LibriSpeech 960~\cite{panayotov2015librispeech}, evaluating WER on test-\{clean, other\} sets. Following~\cite{chang2023exploring}, the discrete codecs are first converted to embeddings by looking up the codebook from the codec models as input to the system. The downstream ASR models and training hyperparameters are aligned with the configurations in~\cite{chang2023exploring}.

\noindent \textbf{Results and Discussion}: The results are shown in Table~\ref{table:codec_task_comparison} (Col.~$2$). The best performance is achieved by Encodec trained on LibriTTS. For AMUSE-based codecs, SoundStream performs the best. While the performance indicating data scaling for codecs is not always applicable to downstream tasks, it also suggests a weak correlation between audio quality metrics and downstream performance.

\subsection{Downstream Integration: TTS}

\noindent \textbf{Setups}: We use LJSpeech~\cite{Ito2017ljspeech} for NAR-TTS and LibriTTS~\cite{zen2019libritts} for AR-TTS. Results are reported in WER by Whisper, UTMOS, and SPK-SIM on the test-clean set.  We follow \cite{shi2024mmm} for the NAR-TTS, including an encoder and multiple decoders to predict 8 stream codes in parallel. For AR-TTS, we follow VALL-E architecture, including an AR decoder and a NAR decoder. The detailed configurations are generally aligned with \cite{wang2023neural}.

\noindent \textbf{Results and Discussion}:
The results are shown in Table~\ref{table:codec_task_comparison} (Col.~$3$-$8$). For NAR-TTS, the model trained with LibriTTS-based SoundStream codec achieves the best intelligibility while the model trained with LibriTTS-based Encodec codec achieves the best UTMOS. For both NAR-TTS and AR-TTS, the performance achieved by LibriTTS-based codecs consistently outperforms their AMUSE counterparts, 
which should be more attributed to the consistency of dataset between codec and the downstream TTS training.

\subsection{Downstream Integration: SSE}

\noindent \textbf{Setups}: We evaluate the model on WSJ0-2Mix with 8kHz sampling rate data by resampling the audio to the 16kHz used by the codec.  Instead of discrete codes, the pre-trained codec encoder and decoder are frozen to serve the SSE system. For the separator and other training configurations, we use identical configurations as in~\cite{yip2024towards}. Models are all trained for 10 epochs using SI-SDR loss. Eight codec streams are used in the experiments.

\noindent \textbf{Results and Discussion}: We report the results in Table~\ref{table:codec_task_comparison} (Col.~$10$-$12$). Different codec types exhibit varying behaviors in SSE performance. While DAC trained on AMUSE has the best PSEQ and STOI scores, Encodec trained on LibriTTS achieves the highest DNSMOS. The data scaling from LibriTTS to AMUSE improves DAC performance but does not necessarily enhance the performance of SoundStream and Encodec.

\subsection{Downstream Integration: SVS}

\noindent \textbf{Setups}: We use Opencpop~\cite{wang2022opencpop} for SVS. Following~\cite{wu2024toksing, tang2024singomd}, we predict 32 codec streams in parallel, with the same hyperparameters as ~\cite{wu2024toksing}. Instead of re-training a vocoder, we use the codec decoder for waveform generation. As LibriTTS models do not consider singing, we only evaluate AMUSE models. We evaluate the system based on MCD, semitone accuracy (SACC), and SingMOS~\cite{tang2024singmos}.

\noindent \textbf{Results and Discussion}: We presents the results in Table~\ref{table:codec_task_comparison}~(Col.~$13$-$15$). Although SoundStream has a worse MCD score than Encodec, both SACC and SingMOS show that it is a better choice overall. The result, however, exhibits mismatches between sound quality measures in Table~\ref{table:main_comparison_large_music}, indicating that SVS performance not only depends on the upper bound set by the codec resynthesis quality but is also related to the difficulties of predicting codec tokens.

\subsection{Downstream Integration: SPK}

\noindent \textbf{Setups}: Our experiments for the SPK task utilize the VoxCeleb1~\cite{nagrani2017voxceleb} corpus. The model is trained using the development set, and performance is reported using the EER on the test set.

\noindent \textbf{Results and Discussion}:
Col.~$9$ of Table~\ref{table:codec_task_comparison} presents the speaker verification performance. 
Codec tokens trained on the larger AMUSE dataset consistently outperform those trained on LibriTTS. 
These results confirm that the discretized codec tokens encapsulate speaker information. 
However, significant overfitting was observed, indicating the need for future research in codec-based speaker verification.

\subsection{Downstream Integration: SSL}

\noindent \textbf{Setups}: We conduct SSL pre-training using LibriSpeech-960 \cite{panayotov2015librispeech}. We perform downstream evaluations on the SSL models using LibriLight Limited \cite{kahnLibriLight} for ASR and SLURP \cite{slurp} for spoken language understanding (SLU). We train variations of HuBERT Base \cite{hsu2021hubert} where the model input uses the quantized neural codec tokens from codec models instead of raw waveforms. We initialize embeddings from 8 codebooks and sum those embeddings across the codebooks. 
After pre-training, we directly fine-tune the models on each downstream task.

\noindent \textbf{Results and Discussion}: The results of the fine-tuned SSL models are shown in Table \ref{table:codec_task_comparison}. We find instability when using quantized codecs as inputs during the pre-training stage, resulting in failed training even after hyperparameter tuning for certain codecs. Aside from downstream performance, another important aspect is the pre-training speed. Our codec-input SSL models only required 200 GPU hours of pre-training, whereas a model using waveform inputs required 280 hours in an equivalent data setting.

\section{Conclusion}\label{sec:concl}


This paper presents ESPnet-Codec, a framework focusing on neural codec training and evaluation in diverse scenarios. Alongside ESPnet-Codec, we introduce VERSA, a unified audio evaluation toolkit designed for comprehensive speech evaluation. Given the recent advancements in neural codecs, we demonstrate the integration of ESPnet-Codec into six downstream tasks. This integration not only showcases the potential of ESPnet-Codec in various applications but also provides a deeper understanding of neural codecs through extensive probing and analysis.

\section{Acknowledgement}

This work used the Bridges2 at PSC and Delta at NCSA through allocations CIS210014 and IRI120008P from the ACCESS program, supported by NSF grants \#2138259, \#2138286, \#2138307, \#2137603, and \#2138296. We also appreciate Wangyou Zhang's advice on VERSA implementation.


\section{References}
{
\printbibliography

@inproceedings{nagrani2017voxceleb,
  title={{VoxCeleb}: A Large-Scale Speaker Identification Dataset},
  author={Nagrani, Arsha and Chung, Joon Son and Zisserman, Andrew},
  booktitle={Proc. Interspeech},
  year={2017},
}

@INPROCEEDINGS{kahnLibriLight,
  author={Kahn, J. and Rivière, M. and Zheng, W. and Kharitonov, E. and Xu, Q. and Mazaré, P.E. and Karadayi, J. and Liptchinsky, V. and Collobert, R. and Fuegen, C. and Likhomanenko, T. and Synnaeve, G. and Joulin, A. and Mohamed, A. and Dupoux, E.},
  booktitle={Proc. ICASSP}, 
  title={{Libri-Light}: A Benchmark for {ASR} with Limited or No Supervision}, 
  year={2020},
  doi={10.1109/ICASSP40776.2020.9052942}
}

@inproceedings{slurp,
    title = "{SLURP}: A Spoken Language Understanding Resource Package",
    author = "Bastianelli, Emanuele  and
      Vanzo, Andrea  and
      Swietojanski, Pawel  and
      Rieser, Verena",
    booktitle = {Proc. EMNLP},
    month = nov,
    year = "2020"
}

@article{hsu2021hubert,
  title={{HuBERT}: Self-supervised speech representation learning by masked prediction of hidden units},
  author={Hsu, Wei-Ning and Bolte, Benjamin and Tsai, Yao-Hung Hubert and Lakhotia, Kushal and Salakhutdinov, Ruslan and Mohamed, Abdelrahman},
  journal={TASLP},
  volume={29},
  pages={3451--3460},
  year={2021},
  publisher={IEEE}
}

@article{chen2022wavlm,
  title={Wav{LM}: Large-scale self-supervised pre-training for full stack speech processing},
  author={Chen, Sanyuan and Wang, Chengyi and Chen, Zhengyang and Wu, Yu and Liu, Shujie and Chen, Zhuo and Li, Jinyu and Kanda, Naoyuki and Yoshioka, Takuya and Xiao, Xiong and others},
  journal={JSTSP},
  volume={16},
  number={6},
  pages={1505--1518},
  year={2022},
  publisher={IEEE}
}

@inproceedings{shi2024multiresolution,
title={Multi-resolution {H}u{BERT}: Multi-resolution Speech Self-Supervised Learning with Masked Unit Prediction},
author={Jiatong Shi and Hirofumi Inaguma and Xutai Ma and Ilia Kulikov and Anna Sun},
booktitle={Proc. ICLR},
year={2024},
}

@inproceedings{chiu2022self,
  title={Self-supervised learning with random-projection quantizer for speech recognition},
  author={Chiu, Chung-Cheng and Qin, James and Zhang, Yu and Yu, Jiahui and Wu, Yonghui},
  booktitle={International Conference on Machine Learning},
  pages={3915--3924},
  year={2022},
  organization={PMLR}
}

@inproceedings{superb,
  author={Shu-Wen Yang and Po-Han Chi and Yung-Sung Chuang and Cheng-I Jeff Lai and Kushal Lakhotia and Yist Y. Lin and Andy T. Liu and Jiatong Shi and Xuankai Chang and Guan-Ting Lin and Tzu-Hsien Huang and Wei-Cheng Tseng and Ko-tik Lee and Da-Rong Liu and Zili Huang and Shuyan Dong and Shang-Wen Li and Shinji Watanabe and Abdelrahman Mohamed and Hung-yi Lee},
  title={{SUPERB: Speech Processing Universal PERformance Benchmark}},
  year=2021,
  booktitle={Proc. Interspeech},
  pages={1194--1198},
  doi={10.21437/Interspeech.2021-1775}
}

@inproceedings{mlsuperb,
  author={Jiatong Shi and Dan Berrebbi and William Chen and En-Pei Hu and Wei-Ping Huang and Ho-Lam Chung and Xuankai Chang and Shang-Wen Li and Abdelrahman Mohamed and Hung-yi Lee and Shinji Watanabe},
  title={{ML-SUPERB: Multilingual Speech Universal PERformance Benchmark}},
  year=2023,
  booktitle={Proc. Interspeech},
  pages={884--888},
  doi={10.21437/Interspeech.2023-1316}
}

@inproceedings{chang2023exploring,
  title={Exploring speech recognition, translation, and understanding with discrete speech units: A comparative study},
  author={Chang, Xuankai and Yan, Brian and Choi, Kwanghee and Jung, Jeeweon and Lu, Yichen and Maiti, Soumi and Sharma, Roshan and Shi, Jiatong and Tian, Jinchuan and Watanabe, Shinji and others},
  booktitle={Proc. ICASSP},
  year={2024}
}

@inproceedings{chang23b_interspeech,
  author={Xuankai Chang and Brian Yan and Yuya Fujita and Takashi Maekaku and Shinji Watanabe},
  title={{Exploration of Efficient End-to-End ASR using Discretized Input from Self-Supervised Learning}},
  year=2023,
  booktitle={Proc. Interspeech},
  pages={1399--1403},
  doi={10.21437/Interspeech.2023-2051}
}

@inproceedings{chang2022distilhubert,
  title={Distill{HuBERT}: Speech representation learning by layer-wise distillation of hidden-unit bert},
  author={Chang, Heng-Jui and Yang, Shu-wen and Lee, Hung-yi},
  booktitle={ICASSP 2022-2022 IEEE International Conference on Acoustics, Speech and Signal Processing (ICASSP)},
  pages={7087--7091},
  year={2022},
  organization={IEEE}
}

@inproceedings{shi2023enhancing,
  title={Enhancing Speech-To-Speech Translation with Multiple {TTS} Targets},
  author={Shi, Jiatong and Tang, Yun and Lee, Ann and Inaguma, Hirofumi and Wang, Changhan and Pino, Juan and Watanabe, Shinji},
  booktitle={ICASSP 2023-2023 IEEE International Conference on Acoustics, Speech and Signal Processing (ICASSP)},
  pages={1--5},
  year={2023},
  organization={IEEE}
}

@inproceedings{yan-etal-2023-espnet,
    title = "{ESP}net-{ST}-v2: Multipurpose Spoken Language Translation Toolkit",
    author = "Yan, Brian  and
      Shi, Jiatong  and
      Tang, Yun  and
      Inaguma, Hirofumi  and
      Peng, Yifan  and
      Dalmia, Siddharth  and
      Pol{\'a}k, Peter  and
      Fernandes, Patrick  and
      Berrebbi, Dan  and
      Hayashi, Tomoki  and
      Zhang, Xiaohui  and
      Ni, Zhaoheng  and
      Hira, Moto  and
      Maiti, Soumi  and
      Pino, Juan  and
      Watanabe, Shinji",
    booktitle = "Proceedings of the 61st Annual Meeting of the Association for Computational Linguistics (Volume 3: System Demonstrations)",
    month = jul,
    year = "2023",
    address = "Toronto, Canada",
    publisher = "Association for Computational Linguistics",
    url = "https://aclanthology.org/2023.acl-demo.38",
    doi = "10.18653/v1/2023.acl-demo.38",
    pages = "400--411",
}

@inproceedings{lee2022direct,
  title={Direct Speech-to-Speech Translation With Discrete Units},
  author={Lee, Ann and Chen, Peng-Jen and Wang, Changhan and Gu, Jiatao and Popuri, Sravya and Ma, Xutai and Polyak, Adam and Adi, Yossi and He, Qing and Tang, Yun and others},
  booktitle={Proceedings of the 60th Annual Meeting of the Association for Computational Linguistics (Volume 1: Long Papers)},
  pages={3327--3339},
  year={2022}
}

@article{zeghidour2021soundstream,
  title={Sound{S}tream: An end-to-end neural audio codec},
  author={Zeghidour, Neil and Luebs, Alejandro and Omran, Ahmed and Skoglund, Jan and Tagliasacchi, Marco},
  journal={TASLP},
  volume={30},
  pages={495--507},
  year={2021},
  publisher={IEEE}
}

@article{yang2023hifi,
  title={Hi{F}i-codec: Group-residual vector quantization for high fidelity audio codec},
  author={Yang, Dongchao and Liu, Songxiang and Huang, Rongjie and Tian, Jinchuan and Weng, Chao and Zou, Yuexian},
  journal={ arXiv:2305.02765},
  year={2023}
}

@article{wang2023neural,
  title={Neural codec language models are zero-shot text to speech synthesizers},
  author={Wang, Chengyi and Chen, Sanyuan and Wu, Yu and Zhang, Ziqiang and Zhou, Long and Liu, Shujie and Chen, Zhuo and Liu, Yanqing and Wang, Huaming and Li, Jinyu and others},
  journal={ arXiv:2301.02111},
  year={2023}
}

@article{wang2023speechx,
  title={Speechx: Neural codec language model as a versatile speech transformer},
  author={Wang, Xiaofei and Thakker, Manthan and Chen, Zhuo and Kanda, Naoyuki and Eskimez, Sefik Emre and Chen, Sanyuan and Tang, Min and Liu, Shujie and Li, Jinyu and Yoshioka, Takuya},
  journal={ arXiv:2308.06873},
  year={2023}
}

@inproceedings{shi2023bridging,
  title={Bridging Speech and Textual Pre-Trained Models With Unsupervised {ASR}},
  author={Shi, Jiatong and Hsu, Chan-Jan and Chung, Holam and Gao, Dongji and Garcia, Paola and Watanabe, Shinji and Lee, Ann and Lee, Hung-yi},
  booktitle={ICASSP 2023-2023 IEEE International Conference on Acoustics, Speech and Signal Processing (ICASSP)},
  pages={1--5},
  year={2023},
  organization={IEEE}
}

@inproceedings{yang2023uniaudio,
  title={Uniaudio: An audio foundation model toward universal audio generation},
  author={Yang, Dongchao and Tian, Jinchuan and Tan, Xu and Huang, Rongjie and Liu, Songxiang and Chang, Xuankai and Shi, Jiatong and Zhao, Sheng and Bian, Jiang and Wu, Xixin and others},
  booktitle={ICML},
  year={2024}
}

@article{wu2024codec,
  title={{Codec-SUPERB}: An In-Depth Analysis of Sound Codec Models},
  author={Wu, Haibin and Chung, Ho-Lam and Lin, Yi-Cheng and Wu, Yuan-Kuei and Chen, Xuanjun and Pai, Yu-Chi and Wang, Hsiu-Hsuan and Chang, Kai-Wei and Liu, Alexander H and Lee, Hung-yi},
  journal={ arXiv:2402.13071},
  year={2024}
}

@inproceedings{panayotov2015librispeech,
  title={Librispeech: an {ASR} corpus based on public domain audio books},
  author={Panayotov, Vassil and Chen, Guoguo and Povey, Daniel and Khudanpur, Sanjeev},
  booktitle={2015 IEEE international conference on acoustics, speech and signal processing (ICASSP)},
  pages={5206--5210},
  year={2015},
  organization={IEEE}
}

@inproceedings{watanabe2018espnet,
  author={Shinji Watanabe and Takaaki Hori and Shigeki Karita and Tomoki Hayashi and Jiro Nishitoba and Yuya Unno and Nelson {Enrique Yalta Soplin} and Jahn Heymann and Matthew Wiesner and Nanxin Chen and Adithya Renduchintala and Tsubasa Ochiai},
  title={{ESPnet}: End-to-End Speech Processing Toolkit},
  year={2018},
  booktitle={Proceedings of Interspeech},
  pages={2207--2211},
  doi={10.21437/Interspeech.2018-1456},
  url={http://dx.doi.org/10.21437/Interspeech.2018-1456}
}

@inproceedings{radford2023robust,
  title={Robust speech recognition via large-scale weak supervision},
  author={Radford, Alec and Kim, Jong Wook and Xu, Tao and Brockman, Greg and McLeavey, Christine and Sutskever, Ilya},
  booktitle={International Conference on Machine Learning},
  pages={28492--28518},
  year={2023},
  organization={PMLR}
}

@inproceedings{hayashi2020espnet,
  title={{ESPnet-TTS}: Unified, reproducible, and integratable open source end-to-end text-to-speech toolkit},
  author={Hayashi, Tomoki and Yamamoto, Ryuichi and Inoue, Katsuki and Yoshimura, Takenori and Watanabe, Shinji and Toda, Tomoki and Takeda, Kazuya and Zhang, Yu and Tan, Xu},
  booktitle={ICASSP 2020-2020 IEEE international conference on acoustics, speech and signal processing (ICASSP)},
  pages={7654--7658},
  year={2020},
  organization={IEEE}
}

@article{hayashi2021espnet2,
  title={{ESP}net2-{TTS}: Extending the edge of {TTS} research},
  author={Hayashi, Tomoki and Yamamoto, Ryuichi and Yoshimura, Takenori and Wu, Peter and Shi, Jiatong and Saeki, Takaaki and Ju, Yooncheol and Yasuda, Yusuke and Takamichi, Shinnosuke and Watanabe, Shinji},
  journal={ arXiv:2110.07840},
  year={2021}
}

@inproceedings{yang2023towards,
  title={Towards Universal Speech Discrete Tokens: A Case Study for {ASR} and {TTS}},
  author={Yang, Yifan and Shen, Feiyu and Du, Chenpeng and Ma, Ziyang and Yu, Kai and Povey, Daniel and Chen, Xie},
  booktitle={Proc. ICASSP},
  year={2024}
}

@article{hayashi2020discretalk,
  title={Discretalk: Text-to-speech as a machine translation problem},
  author={Hayashi, Tomoki and others},
  journal={ arXiv:2005.05525},
  year={2020}
}

@article{defossez2022high,
  title={High Fidelity Neural Audio Compression},
  author={D{\'e}fossez, Alexandre and Copet, Jade and Synnaeve, Gabriel and Adi, Yossi},
  journal={TMLR},
  year={2023}
}

@inproceedings{mustafa2021stylemelgan,
  title={Style{M}el{GAN}: An efficient high-fidelity adversarial vocoder with temporal adaptive normalization},
  author={Mustafa, Ahmed and Pia, Nicola and Fuchs, Guillaume},
  booktitle={ICASSP 2021-2021 IEEE International Conference on Acoustics, Speech and Signal Processing (ICASSP)},
  pages={6034--6038},
  year={2021},
}

@article{kumar2024high,
  title={High-fidelity audio compression with improved {RVQGAN}},
  author={Kumar, Rithesh and Seetharaman, Prem and Luebs, Alejandro and Kumar, Ishaan and Kumar, Kundan},
  journal={Proc. NeurIPS},
  volume={36},
  year={2024}
}

@inproceedings{copet2023simple,
    title={Simple and Controllable Music Generation},
    author={Jade Copet and Felix Kreuk and Itai Gat and Tal Remez and David Kant and Gabriel Synnaeve and Yossi Adi and Alexandre Défossez},
    booktitle={Proc. NeurIPS},
    year={2023},
}

@inproceedings{du2024funcodec,
  title={Fun{C}odec: A fundamental, reproducible and integrable open-source toolkit for neural speech codec},
  author={Du, Zhihao and Zhang, Shiliang and Hu, Kai and Zheng, Siqi},
  booktitle={Proc. ICASSP},
  pages={591--595},
  year={2024},
  organization={IEEE}
}

@inproceedings{
zhang2024speechtokenizer,
title={Speech{T}okenizer: Unified Speech Tokenizer for Speech Language Models},
author={Xin Zhang and Dong Zhang and Shimin Li and Yaqian Zhou and Xipeng Qiu},
booktitle={Proc. ICLR},
year={2024},
url={https://openreview.net/forum?id=AF9Q8Vip84}
}

@article{zhang2023amphion,
  title={Amphion: An Open-Source Audio, Music and Speech Generation Toolkit},
  author={Zhang, Xueyao and Xue, Liumeng and Wang, Yuancheng and Gu, Yicheng and Chen, Xi and Fang, Zihao and Chen, Haopeng and Zou, Lexiao and Wang, Chaoren and Han, Jun and others},
  journal={ arXiv:2312.09911},
  year={2023}
}

@inproceedings{li2021real,
  title={Real-time speech frequency bandwidth extension},
  author={Li, Yunpeng and Tagliasacchi, Marco and Rybakov, Oleg and Ungureanu, Victor and Roblek, Dominik},
  booktitle={ICASSP 2021-2021 IEEE International Conference on Acoustics, Speech and Signal Processing (ICASSP)},
  pages={691--695},
  year={2021},
}

@article{razavi2019generating,
  title={Generating diverse high-fidelity images with {VQ-VAE}-2},
  author={Razavi, Ali and Van den Oord, Aaron and Vinyals, Oriol},
  journal={Proc. NeurIPS},
  volume={32},
  year={2019}
}

@article{kong2020hifi,
  title={{HiFi-GAN}: Generative adversarial networks for efficient and high fidelity speech synthesis},
  author={Kong, Jungil and Kim, Jaehyeon and Bae, Jaekyoung},
  journal={Proc. NeurIPS},
  volume={33},
  pages={17022--17033},
  year={2020}
}

@inproceedings{shi2022muskits,
  title={Muskits: an End-to-End Music Processing Toolkit for Singing Voice Synthesis},
  author={Shi, Jiatong and Guo, Shuai and Qian, Tao and others},
  booktitle={Interspeech},
  year={2022}
}

@inproceedings{jung2024espnet,
  title={{ESPnet-SPK}: full pipeline speaker embedding toolkit with reproducible recipes, self-supervised front-ends, and off-the-shelf models},
  author={Jung, Jee-weon and Zhang, Wangyou and Shi, Jiatong and Aldeneh, Zakaria and Higuchi, Takuya and Theobald, Barry-John and Abdelaziz, Ahmed Hussen and Watanabe, Shinji},
  booktitle={Interspeech},
  year={2024}
}

@inproceedings{li2021espnet,
  title={{ESPnet-SE}: End-to-end speech enhancement and separation toolkit designed for ASR integration},
  author={Li, Chenda and Shi, Jing and Zhang, Wangyou and Subramanian, Aswin Shanmugam and Chang, Xuankai and Kamo, Naoyuki and Hira, Moto and Hayashi, Tomoki and Boeddeker, Christoph and Chen, Zhuo and others},
  booktitle={2021 IEEE Spoken Language Technology Workshop (SLT)},
  pages={785--792},
  year={2021},
  organization={IEEE}
}

@inproceedings{lu2022espnet,
  title={{ESPnet-SE}++: Speech Enhancement for Robust Speech Recognition, Translation, and Understanding},
  author={Lu, Yen Ju and Chang, Xuankai and Li, Chenda and Zhang, Wangyou and Cornell, Samuele and Ni, Zhaoheng and Masuyama, Yoshiki and Yan, Brian and Scheibler, Robin and Wang, Zhong Qiu and others},
  booktitle={Proc. Interspeech},
  volume={2022},
  pages={5458--5462},
  year={2022}
}

@inproceedings{chen2023reducing,
  title={Reducing Barriers to Self-Supervised Learning: {HuBERT} Pre-training with Academic Compute},
  author={Chen, William and Chang, Xuankai and Peng, Yifan and Ni, Zhaoheng and Maiti, Soumi and Watanabe, Shinji},
  booktitle={Interspeech},
  volume={2023},
  pages={4404--4408},
  year={2023}
}

@inproceedings{wu2023audiodec,
  title={Audiodec: An open-source streaming high-fidelity neural audio codec},
  author={Wu, Yi-Chiao and Gebru, Israel D and Markovi{\'c}, Dejan and Richard, Alexander},
  booktitle={ICASSP 2023-2023 IEEE International Conference on Acoustics, Speech and Signal Processing (ICASSP)},
  pages={1--5},
  year={2023},
  organization={IEEE}
}

@inproceedings{junaturalspeech,
  title={NaturalSpeech 3: Zero-Shot Speech Synthesis with Factorized Codec and Diffusion Models},
  author={Ju, Zeqian and Wang, Yuancheng and Shen, Kai and Tan, Xu and Xin, Detai and Yang, Dongchao and Liu, Eric and Leng, Yichong and Song, Kaitao and Tang, Siliang and others},
  booktitle={ICML},
  year={2024}
}

@inproceedings{zhang2023speechgpt,
  title={Speech{GPT}: Empowering Large Language Models with Intrinsic Cross-Modal Conversational Abilities},
  author={Zhang, Dong and Li, Shimin and Zhang, Xin and Zhan, Jun and Wang, Pengyu and Zhou, Yaqian and Qiu, Xipeng},
  booktitle={Proc. EMNLP},
  year={2023}
}

@article{yang2024simplespeech,
  title={SimpleSpeech: Towards Simple and Efficient Text-to-Speech with Scalar Latent Transformer Diffusion Models},
  author={Yang, Dongchao and Wang, Dingdong and Guo, Haohan and Chen, Xueyuan and Wu, Xixin and Meng, Helen},
  journal={ arXiv:2406.02328},
  year={2024}
}

@article{liu2023high,
  title={A High Fidelity and Low Complexity Neural Audio Coding},
  author={Liu, Wenzhe and Xiao, Wei and Wang, Meng and Yang, Shan and Shi, Yupeng and Kang, Yuyong and Su, Dan and Shang, Shidong and Yu, Dong},
  journal={ arXiv:2310.10992},
  year={2023}
}

@inproceedings{le2019sdr,
  title={{SDR}--half-baked or well done?},
  author={Le Roux, Jonathan and Wisdom, Scott and Erdogan, Hakan and Hershey, John R},
  booktitle={ICASSP 2019-2019 IEEE International Conference on Acoustics, Speech and Signal Processing (ICASSP)},
  pages={626--630},
  year={2019},
  organization={IEEE}
}

@inproceedings{boeddeker2021convolutive,
  title={Convolutive transfer function invariant {SDR} training criteria for multi-channel reverberant speech separation},
  author={Boeddeker, Christoph and Zhang, Wangyou and Nakatani, Tomohiro and Kinoshita, Keisuke and Ochiai, Tsubasa and Delcroix, Marc and Kamo, Naoyuki and Qian, Yanmin and Haeb-Umbach, Reinhold},
  booktitle={ICASSP 2021-2021 IEEE International Conference on Acoustics, Speech and Signal Processing (ICASSP)},
  pages={8428--8432},
  year={2021},
  organization={IEEE}
}

@inproceedings{rix2001perceptual,
  title={Perceptual evaluation of speech quality ({PESQ})-a new method for speech quality assessment of telephone networks and codecs},
  author={Rix, Antony W and Beerends, John G and Hollier, Michael P and Hekstra, Andries P},
  booktitle={Proc. ICASSP},
  volume={2},
  pages={749--752},
  year={2001},
  organization={IEEE}
}

@article{taal2011algorithm,
  title={An algorithm for intelligibility prediction of time--frequency weighted noisy speech},
  author={Taal, Cees H and Hendriks, Richard C and Heusdens, Richard and Jensen, Jesper},
  journal={TASLP},
  volume={19},
  number={7},
  pages={2125--2136},
  year={2011},
  publisher={IEEE}
}

@article{hines2015visqol,
  title={{ViSQOL}: an objective speech quality model},
  author={Hines, Andrew and Skoglund, Jan and Kokaram, Anil C and Harte, Naomi},
  journal={JASM},
  volume={2015},
  pages={1--18},
  year={2015},
  publisher={Springer}
}

@article{saeki2024speechbertscore,
  title={Speech{BERTS}core: Reference-Aware Automatic Evaluation of Speech Generation Leveraging NLP Evaluation Metrics},
  author={Saeki, Takaaki and Maiti, Soumi and Takamichi, Shinnosuke and Watanabe, Shinji and Saruwatari, Hiroshi},
  journal={ arXiv:2401.16812},
  year={2024}
}

@inproceedings{reddy2021dnsmos,
  title={{DNSMOS}: A non-intrusive perceptual objective speech quality metric to evaluate noise suppressors},
  author={Reddy, Chandan KA and Gopal, Vishak and Cutler, Ross},
  booktitle={ICASSP 2021-2021 IEEE International Conference on Acoustics, Speech and Signal Processing (ICASSP)},
  pages={6493--6497},
  year={2021},
  organization={IEEE}
}

@article{saeki2022utmos,
  title={{UTMOS}: UTokyo-SaruLab System for VoiceMOS Challenge 2022},
  author={Saeki, Takaaki and Xin, Detai and Nakata, Wataru and Koriyama, Tomoki and Takamichi, Shinnosuke and Saruwatari, Hiroshi},
  journal={Interspeech},
  year={2022},
  publisher={ISCA}
}

@article{dienerplcmos,
  title={{PLCMOS}--a data-driven non-intrusive metric for the evaluation of packet loss concealment algorithms},
  author={Diener, Lorenz and Purin, Marju and Sootla, Sten and Saabas, Ando and Aichner, Robert and Cutler, Ross},
  journal={Interspeech},
  year={2022},
  publisher={ISCA}
}

@inproceedings{zen2019libritts,
  title={Libri{TTS}: A Corpus Derived from LibriSpeech for Text-to-Speech},
  author={Zen, Heiga and Dang, Viet and Clark, Rob and Zhang, Yu and Weiss, Ron J and Jia, Ye and Chen, Zhifeng and Wu, Yonghui},
  year={2019},
  booktitle={Proc. Interspeech},
}

@inproceedings{shi2021aishell,
  title={{AISHELL-3}: A Multi-Speaker Mandarin TTS Corpus},
  author={Shi, Yao and Bu, Hui and Xu, Xin and Zhang, Shaoji and Li, Ming},
  year={2021},
  booktitle={Proc. Interspeech},
}

@inproceedings{shi2021highland,
  title={Highland {P}uebla {N}ahuatl speech translation corpus for endangered language documentation},
  author={Shi, Jiatong and Amith, Jonathan D and Chang, Xuankai and Dalmia, Siddharth and Yan, Brian and Watanabe, Shinji},
  booktitle={Proc. AmericasNLP},
  pages={53--63},
  year={2021}
}

@inproceedings{shi2021leveraging,
  title={Leveraging End-to-End {ASR} for Endangered Language Documentation: An Empirical Study on {Y}ol{\'o}xochitl {M}ixtec},
  author={Shi, Jiatong and Amith, Jonathan D and Garc{\'\i}a, Rey Castillo and Sierra, Esteban Guadalupe and Duh, Kevin and Watanabe, Shinji},
  booktitle={Proc. EACL},
  pages={1134--1145},
  year={2021}
}

@inproceedings{huang2021multi,
  title={Multi-singer: Fast multi-singer singing voice vocoder with a large-scale corpus},
  author={Huang, Rongjie and Chen, Feiyang and Ren, Yi and Liu, Jinglin and Cui, Chenye and Zhao, Zhou},
  booktitle={Proc. ACMMM},
  pages={3945--3954},
  year={2021}
}

@inproceedings{dai2023singstyle111,
  title={Singstyle111: A multilingual singing dataset with style transfer},
  author={Dai, Shuqi and Chen, Siqi and Wu, Yuxuan and Diao, Ruxin and Huang, Roy and Dannenberg, Roger B},
  booktitle={Proc. ISMIR},
  volume={1},
  pages={4--2},
  year={2023}
}

@article{zhang2022m4singer,
  title={M4singer: A multi-style, multi-singer and musical score provided mandarin singing corpus},
  author={Zhang, Lichao and Li, Ruiqi and Wang, Shoutong and Deng, Liqun and Liu, Jinglin and Ren, Yi and He, Jinzheng and Huang, Rongjie and Zhu, Jieming and Chen, Xiao and others},
  journal={Proc. NeurIPS},
  volume={35},
  pages={6914--6926},
  year={2022}
}

@article{ogawa2021tohoku,
  title={Tohoku {K}iritan singing database: A singing database for statistical parametric singing synthesis using Japanese pop songs},
  author={Ogawa, Itsuki and Morise, Masanori},
  journal={AST},
  volume={42},
  number={3},
  pages={140--145},
  year={2021},
  publisher={Acoustical Society of Japan}
}

@inproceedings{wang2022opencpop,
  title={Opencpop: A High-Quality Open Source Chinese Popular Song Corpus for Singing Voice Synthesis},
  author={Wang, Yu and Wang, Xinsheng and Zhu, Pengcheng and Wu, Jie and Li, Hanzhao and Xue, Heyang and Zhang, Yongmao and Xie, Lei and Bi, Mengxiao},
  year={2022},
  booktitle={Proc. Interspeech},
}

@inproceedings{shi2024singing,
  title={Singing Voice Data Scaling-up: An Introduction to {ACE-Opencpop and ACE-KiSing}},
  author={Shi, Jiatong and Lin, Yueqian and Bai, Xinyi and Zhang, Keyi and Wu, Yuning and Tang, Yuxun and Yu, Yifeng and Jin, Qin and Watanabe, Shinji},
  booktitle={Proc. Interspeech},
  year={2024}
}

@inproceedings{koguchi2020pjs,
  title={{PJS}: Phoneme-balanced Japanese singing-voice corpus},
  author={Koguchi, Junya and Takamichi, Shinnosuke and Morise, Masanori},
  booktitle={Proc. APSIPA ASC},
  pages={487--491},
  year={2020},
  organization={IEEE}
}

@article{takamichi2020jsut,
  title={{JSUT and JVS}: Free Japanese voice corpora for accelerating speech synthesis research},
  author={Takamichi, Shinnosuke and Sonobe, Ryosuke and Mitsui, Kentaro and Saito, Yuki and Koriyama, Tomoki and Tanji, Naoko and Saruwatari, Hiroshi},
  journal={AST},
  volume={41},
  number={5},
  pages={761--768},
  year={2020},
}

@inproceedings{yamamoto2020parallel,
  title={Parallel {WaveGAN}: A fast waveform generation model based on generative adversarial networks with multi-resolution spectrogram},
  author={Yamamoto, Ryuichi and Song, Eunwoo and Kim, Jae-Min},
  booktitle={ICASSP 2020-2020 IEEE International Conference on Acoustics, Speech and Signal Processing (ICASSP)},
  pages={6199--6203},
  year={2020},
  organization={IEEE}
}

@inproceedings{yip2024towards,
  title={Towards Audio Codec-based Speech Separation},
  author={Yip, Jia Qi and Zhao, Shengkui and Ng, Dianwen and Chng, Eng Siong and Ma, Bin},
  booktitle={Proc. Interspeech},
  year={2024}
}

@inproceedings{chen2024loss,
  title={Loss Masking Is Not Needed In Decoder-Only Transformer For Discrete-Token-Based {ASR}},
  author={Chen, Qian and Wang, Wen and Zhang, Qinglin and Zheng, Siqi and Zhang, Shiliang and Deng, Chong and Ma, Yukun and Yu, Hai and Liu, Jiaqing and Zhang, Chong},
  booktitle={ICASSP 2024-2024 IEEE International Conference on Acoustics, Speech and Signal Processing (ICASSP)},
  pages={11056--11060},
  year={2024},
  organization={IEEE}
}

@inproceedings{chang2024interspeech,
  title={The {I}nterspeech 2024 Challenge on Speech Processing Using Discrete Units},
  author={Chang, Xuankai and Shi, Jiatong and Tian, Jinchuan and Wu, Yuning and Tang, Yuxun and Wu, Yihan and Watanabe, Shinji and Adi, Yossi and Chen, Xie and Jin, Qin},
  booktitle={Proc. Interspeech},
  year={2024}
}

@article{ye2023asq,
  title={{ASQ}: An Ultra-Low Bit Rate {ASR}-Oriented Speech Quantization Method},
  author={Ye, Lingxuan and Gao, Changfeng and Cheng, Gaofeng and Luo, Liuping and Zhao, Qingwei},
  journal={IEEE Signal Processing Letters},
  year={2023},
  publisher={IEEE}
}

@inproceedings{shi2024mmm,
  title={{MMM}: Multi-Layer Multi-Residual Multi-Stream Discrete Speech Representation from Self-supervised Learning Model},
  author={Shi, Jiatong and Ma, Xutai and Inaguma, Hirofumi and Sun, Anna and Watanabe, Shinji},
  booktitle={Proc. Interspeech},
  year={2024}
}

@article{wang2023viola,
  title={Viola: Unified codec language models for speech recognition, synthesis, and translation},
  author={Wang, Tianrui and Zhou, Long and Zhang, Ziqiang and Wu, Yu and Liu, Shujie and Gaur, Yashesh and Chen, Zhuo and Li, Jinyu and Wei, Furu},
  journal={ arXiv:2305.16107},
  year={2023}
}

@article{gupta2024exploring,
  title={Exploring the limits of decoder-only models trained on public speech recognition corpora},
  author={Gupta, Ankit and Saon, George and Kingsbury, Brian},
  journal={ arXiv:2402.00235},
  year={2024}
}

@inproceedings{ren2020fastspeech,
  title={Fast{S}peech 2: Fast and High-Quality End-to-End Text to Speech},
  author={Ren, Yi and Hu, Chenxu and Tan, Xu and Qin, Tao and Zhao, Sheng and Zhao, Zhou and Liu, Tie-Yan},
  booktitle={Proc. ICLR},
  year={2020}
}

@inproceedings{kim2021conditional,
  title={Conditional variational autoencoder with adversarial learning for end-to-end text-to-speech},
  author={Kim, Jaehyeon and Kong, Jungil and Son, Juhee},
  booktitle={Proc. ICML},
  pages={5530--5540},
  year={2021},
  organization={PMLR}
}

@inproceedings{shen2018natural,
  title={Natural {TTS} synthesis by conditioning wavenet on mel spectrogram predictions},
  author={Shen, Jonathan and Pang, Ruoming and Weiss, Ron J and Schuster, Mike and Jaitly, Navdeep and Yang, Zongheng and Chen, Zhifeng and Zhang, Yu and Wang, Yuxuan and Skerrv-Ryan, Rj and others},
  booktitle={2018 IEEE international conference on acoustics, speech and signal processing (ICASSP)},
  pages={4779--4783},
  year={2018},
  organization={IEEE}
}

@article{anastassiou2024seed,
  title={Seed-{TTS}: A Family of High-Quality Versatile Speech Generation Models},
  author={Anastassiou, Philip and Chen, Jiawei and Chen, Jitong and Chen, Yuanzhe and Chen, Zhuo and Chen, Ziyi and Cong, Jian and Deng, Lelai and Ding, Chuang and Gao, Lu and others},
  journal={ arXiv:2406.02430},
  year={2024}
}

@inproceedings{wu2024toksing,
  title={TokSing: Singing Voice Synthesis based on Discrete Tokens},
  author={Wu, Yuning and Shi, Jiatong and Tang, Yuxun and Yang, Shan and Jin, Qin and others},
  booktitle={Proc. Interspeech},
  year={2024}
}

@inproceedings{tang2024singomd,
  title={Sing{OMD}: Singing Oriented Multi-resolution Discrete Representation Construction from Speech Models},
  author={Tang, Yuxun and Wu, Yuning and Shi, Jiatong and Jin, Qin},
  booktitle={Proc. Interspeech},
  year={2024}
}

@inproceedings{zhang23e_interspeech,
  author={Yongmao Zhang and Heyang Xue and Hanzhao Li and Lei Xie and Tingwei Guo and Ruixiong Zhang and Caixia Gong},
  title={{VISinger2: High-Fidelity End-to-End Singing Voice Synthesis Enhanced by Digital Signal Processing Synthesizer}},
  year=2023,
  booktitle={Proc. Interspeech},
  pages={4444--4448},
  doi={10.21437/Interspeech.2023-391},
  issn={2958-1796}
}

@inproceedings{shi2021sequence,
  title={Sequence-to-sequence singing voice synthesis with perceptual entropy loss},
  author={Shi, Jiatong and Guo, Shuai and Huo, Nan and Zhang, Yuekai and Jin, Qin},
  booktitle={ICASSP 2021-2021 IEEE International Conference on Acoustics, Speech and Signal Processing (ICASSP)},
  pages={76--80},
  year={2021},
  organization={IEEE}
}

@article{Wang2022TFGridNetIF,
  title={{TF-GridNet}: Integrating Full- and Sub-Band Modeling for Speech Separation},
  author={Zhongqiu Wang and Samuele Cornell and Shukjae Choi and Younglo Lee and Byeonghak Kim and Shinji Watanabe},
  journal={TASLP},
  year={2022},
  volume={31},
  pages={3221-3236},
}

@article{Subakan2022ExploringSM,
  title={Exploring Self-Attention Mechanisms for Speech Separation},
  author={Cem Subakan and Mirco Ravanelli and Samuele Cornell and François Grondin and Mirko Bronzi},
  journal={TASLP},
  year={2022},
  volume={31},
  pages={2169-2180},
}

@inproceedings{Zhao2023MossFormer2CT,
  title={{MossFormer2}: Combining Transformer and RNN-Free Recurrent Network for Enhanced Time-Domain Monaural Speech Separation},
  author={Shengkui Zhao and Yukun Ma and Chongjia Ni and Chong Zhang and Hao Wang and Trung Hieu Nguyen and Kun Zhou and Jia Qi Yip and Dianwen Ng and Bin Ma},
  booktitle={ICASSP 2024-2024 IEEE International Conference on Acoustics, Speech and Signal Processing (ICASSP)},
  year={2024},
  organization={IEEE}
}

@inproceedings{Yip2023SPGMPL,
  title={{SPGM}: Prioritizing Local Features for enhanced speech separation performance},
  author={Jia Qi Yip and Shengkui Zhao and Yukun Ma and Chongjia Ni and Chong Zhang and Hao Wang and Trung Hieu Nguyen and Kun Zhou and Dianwen Ng and Eng Siong Chng and Binchao Ma},
  booktitle={ICASSP 2024-2024 IEEE International Conference on Acoustics, Speech and Signal Processing (ICASSP)},
  year={2024},
  organization={IEEE}
}

@INPROCEEDINGS{Song2022slt,
  author={Song, Hyungchan and Chen, Sanyuan and Chen, Zhuo and Wu, Yu and Yoshioka, Takuya and Tang, Min and Shin, Jong Won and Liu, Shujie},
  booktitle={2022 IEEE Spoken Language Technology Workshop (SLT)}, 
  title={Exploring {WavLM} on Speech Enhancement}, 
  year={2023},
  volume={},
  number={},
  pages={451--457},
}

@ARTICLE{Richter2023,
  author={Richter, Julius and Welker, Simon and Lemercier, Jean-Marie and Lay, Bunlong and Gerkmann, Timo},
  journal={TASLP}, 
  title={Speech Enhancement and Dereverberation With Diffusion-Based Generative Models}, 
  year={2023},
  volume={31},
  number={},
  pages={2351--2364},
}

@inproceedings{puvvada2024discrete,
  title={Discrete Audio Representation as an Alternative to Mel-Spectrograms for Speaker and Speech Recognition},
  author={Puvvada, Krishna C and Koluguri, Nithin Rao and Dhawan, Kunal and Balam, Jagadeesh and Ginsburg, Boris},
  booktitle={ICASSP 2024-2024 IEEE International Conference on Acoustics, Speech and Signal Processing (ICASSP)},
  pages={12111--12115},
  year={2024},
  organization={IEEE}
}

@inproceedings{jung2022pushing,
  title={Pushing the limits of raw waveform speaker recognition},
  author={Jung, Jee-weon and Kim, Youjin and Heo, Hee-Soo and Lee, Bong-Jin and Kwon, Youngki and Chung, Joon Son},
  booktitle={Proc. Interspeech},
  pages={2228--2232},
  year={2022},
}

@inproceedings{desplanques2020ecapa,
  title={{ECAPA-TDNN}: Emphasized Channel Attention, Propagation and Aggregation in {TDNN} based speaker verification},
  author={Desplanques, Brecht and Thienpondt, Jenthe and Demuynck, Kris},
  booktitle={Proc. Interspeech},
  pages={3830--3834},
  year={2020},
  organization={International Speech Communication Association (ISCA)}
}

@inproceedings{yang2023fast,
  title={Fast-{H}ubert: an Efficient Training Framework for Self-Supervised Speech Representation Learning},
  author={Yang, Guanrou and Ma, Ziyang and Zheng, Zhisheng and Song, Yakun and Niu, Zhikang and Chen, Xie},
  booktitle={2023 IEEE Automatic Speech Recognition and Understanding Workshop (ASRU)},
  pages={1--7},
  year={2023},
  organization={IEEE}
}

@inproceedings{lugo2024towards,
  title={Towards efficient self-supervised representation learning in speech processing},
  author={Lugo, Luis and Vielzeuf, Valentin},
  booktitle={Proc. EACL},
  pages={340--346},
  year={2024}
}

@inproceedings{parcollet2023efficiency,
  title={On the (in) efficiency of acoustic feature extractors for self-supervised speech representation learning},
  author={Parcollet, Titouan and Zhang, Shucong and van Dalen, Rogier and Ramos, Alberto Gil CP and Bhattacharya, Sourav},
  booktitle={Proc. Interspeech},
  year={2023}
}

@inproceedings{lin2023melhubert,
  title={MelHuBERT: A simplified HuBERT on Mel spectrograms},
  author={Lin, Tzu-Quan and Lee, Hung-yi and Tang, Hao},
  booktitle={2023 IEEE Automatic Speech Recognition and Understanding Workshop (ASRU)},
  pages={1--8},
  year={2023},
  organization={IEEE}
}

@inproceedings{liu2022diffsinger,
  title={Diffsinger: Singing voice synthesis via shallow diffusion mechanism},
  author={Liu, Jinglin and Li, Chengxi and Ren, Yi and Chen, Feiyang and Zhao, Zhou},
  booktitle={Proc. AAAI},
  volume={36},
  number={10},
  pages={11020--11028},
  year={2022}
}

@article{tang2024singmos,
  title={{SingMOS}: An extensive Open-Source Singing Voice Dataset for MOS Prediction},
  author={Tang, Yuxun and Shi, Jiatong and Wu, Yuning and Jin, Qin},
  journal={ arXiv:2406.10911},
  year={2024}
}

@misc{Ito2017ljspeech,
  author       = {Keith Ito and Linda Johnson},
  title        = {The LJ Speech Dataset},
  howpublished = {\url{https://keithito.com/LJ-Speech-Dataset/}},
  year         = 2017
}

@misc{amith_yoloxochitl_mixtec,
  author = {Amith, Jonathan D. and Castillo Castillo, Rey},
  title = {Audio corpus of Yoloxóchitl Mixtec with accompanying time-coded transcriptons in ELAN},
  note = {n.d.},
url={https://www.openslr.org/89/}
}

@misc{amith_audio_corpus_sierra,
  author = {Amith, Jonathan D. and Domínguez Alcántara, Amelia and Salazar Osollo, Hermelindo and Salgado Castañeda, Ceferino and Gorostiza Salazar, Eleuterio},
  title = {Audio corpus of Sierra Nororiental and Sierra Norte de Puebla Nahuat(l) with accompanying time-code transcriptions in ELAN},
  note = {n.d.},
url={https://www.openslr.org/92/}
}

@misc{amith_totonac,
  author = {Amith, Jonathan D. and López Francisco, Osbel},
  title = {Audio corpus of Totonac recordings from northern Puebla and adjacent areas of Veracruz},
  note = {n.d.},
  url = {https://www.openslr.org/107/}
}
}


\end{document}